\documentclass{ieeeaccess}
\usepackage{cite}
\usepackage{amsmath,amssymb,amsfonts,subfigure}
\usepackage{algorithmic}
\usepackage{graphicx}
\usepackage{textcomp}
\usepackage{epsfig,epstopdf}
\usepackage{booktabs}
\usepackage{multirow}
\usepackage{soul}
\usepackage{nomencl}
\makeindex
    \makenomenclature
\soulregister\cite7 
\soulregister\ref7 
\def\BibTeX{{\text B\kern-.05em{\sc i\kern-.025em b}\kern-.08em
    T\kern-.1667em\lower.7ex\hbox{E}\kern-.125emX}}

\begin{document}

\history{Date of publication xxxx 00, 0000, date of current version xxxx 00, 0000.}
\doi{10.1109/ACCESS.2020.DOI}

\title{A 3D Non-Stationary GBSM for Vehicular Visible Light Communication MISO Channels}
\author{\uppercase{Ahmed~Al-Kinani}\authorrefmark{1},
\uppercase{Cheng-Xiang~Wang}\authorrefmark{2,3,4},~\IEEEmembership{Fellow, IEEE},
\uppercase{Qiuming Zhu}\authorrefmark{5}, \IEEEmembership{Member, IEEE},
\uppercase{Yu Fu}\authorrefmark{4},
\uppercase{el-Hadi~M.Aggoune}\authorrefmark{6}, \IEEEmembership{Life Senior Member, IEEE},
\uppercase{Ahmed Talib}\authorrefmark{7}, AND
\uppercase{Nidaa Al-Hasaani}\authorrefmark{7}}
\address[1]{Cellular Asset Management Services Ltd., Surrey TW20 8HE, U.K.}
\address[2]{National Mobile Communications Research Laboratory, School of Information Science and Engineering, Southeast University, Nanjing, 210096, China}
\address[3]{Purple Mountain Laboratories, Nanjing 211111, China}
\address[4]{Institute of Sensors, Signals and Systems, School of Engineering \& Physical Sciences, Heriot-Watt University, Edinburgh EH14 4AS, U.K.}
\address[5]{College of Electronic and Information Engineering, Nanjing University of Aeronautics and Astronautics, Nanjing 211106, China}
\address[6]{Sensor Networks and Cellular Systems Research Center, University of Tabuk, Tabuk 47315/4031, Saudi Arabia}
\address[7]{Ministry of Communications of Iraq, Baghdad 10013, Iraq}
\tfootnote{This work was supported by the National Key R\&D Program of China under Grant 2018YFB1801101, the National Natural Science Foundation of China (NSFC) under Grant 61960206006 and 61871035, the Frontiers Science Center for Mobile Information Communication and Security, the High Level Innovation and Entrepreneurial Research Team Program in Jiangsu, the High Level Innovation and Entrepreneurial Talent Introduction Program in Jiangsu, the Research Fund of National Mobile Communications Research Laboratory, Southeast University, under Grant 2020B01, the Fundamental Research Funds for the Central Universities under Grant 2242020R30001, the Huawei Cooperation Project, the EU H2020 RISE TESTBED2 project under Grant 872172, the Ministry of Communications of Iraq under Grant 821978, and the Sensor Networks and Cellular Systems (SNCS) Research Center, University of Tabuk under Grant 1440-503.}

\markboth
{A. Al-Kinani \headeretal: A 3D Non-Stationary GBSM for Vehicular Visible Light Communication MISO Channels\hspace{-0.8cm}}
{\hspace{-0.8cm}A. Al-Kinani \headeretal: A 3D Non-Stationary GBSM for Vehicular Visible Light Communication MISO Channels}

\corresp{Corresponding author: Cheng-Xiang~Wang (chxwang@seu.edu.cn).}

\begin{abstract}
The potential of using visible light communication (VLC) technologies for vehicular communication networks has recently attracted much attention. The underlying VLC channels, as a foundation for the proper design and optimization of vehicular VLC communication systems, have not yet been sufficiently investigated. Vehicular VLC link impairments can have a significant impact on the system performance and capacity. Such impairments include the optical wireless channel distortion and background noise. This paper proposes a novel three-dimensional (3D) regular-shaped geometry-based stochastic model (RS-GBSM) for vehicular VLC multiple-input single-output (MISO) channels. The proposed 3D RS-GBSM combines a two-sphere model and an elliptic-cylinder model. Both the line-of-sight (LoS) and single-bounced (SB) components are considered. The proposed model jointly considers the azimuth and elevation angles by using von-Mises-Fisher (VMF) distribution. Based on the proposed model, the relationship between the communication range and the received optical power is analyzed and validated by simulations. The impact of the elevation angle in the 3D model on the received optical power is investigated by comparing with the received optical power of the corresponding two-dimensional (2D) model. Furthermore, the background noise is also modeled to evaluate the system's signal-to-noise ratio (SNR).
\end{abstract}


\titlepgskip=-15pt

\maketitle

\nomenclature{$N$}{Number of clusters.}%
\nomenclature{$M$}{Number of rays within each cluster.}%
\nomenclature{$U$}{Number of receiver (Rx) antenna elements.}%
\nomenclature{$S$}{Number of transmitter (Tx) antenna elements.}%
\nomenclature{$\rho$}{Polarization.}
\nomenclature{$L$}{Mean value of the number of newly generated clusters.}%
\nomenclature{$T$}{The number of time samples.}%
\nomenclature{$s_{\text {DS}}$}{Log-normal distributed RV of delay spread (DS).}%
\nomenclature{$s_{\text {ASA}}$}{Log-normal distributed random variable (RV) of angle spread of arrival (ASA).}%
\nomenclature{$s_{\text {ASD}}$}{Log-normal distributed RV of angle spread of departure (ASD).}%
\nomenclature{$s_{\text {SF}}$}{Log-normal distributed RV of shadow fading (SF).}%
\nomenclature{$s_{{K}}$}{Log-normal distributed RV of Rician K-factor.}%
\nomenclature{$P_n$}{Power of the $n$-th  cluster.}%
\nomenclature{$\tau_n$}{Normalized delay of the $n$-th ($n=1, \cdots, N$) cluster.}%
\nomenclature{${\Phi_{m,n}}$}{Random initial phases related to the $m$-th ray within the $n$-th cluster.}%
\nomenclature{${\varphi _{n,m}}$}{Angle of arrival (AoA) related to the $m$-th ray within the $n$-th cluster.}%
\nomenclature{${\phi _{n,m}}$}{Angle of departure (AoD) related to the $m$-th ($m=1, \cdots, M$) ray within the $n$-th cluster.}%
\nomenclature{${\upsilon _{n,m}}$}{Doppler frequency component related to the $m$-th ray within the $n$-th cluster.}%
\nomenclature{$v$, ${\theta _v}$}{Speed and mobile direction of MS, respectively.}%
\nomenclature{$v_c$, ${\theta _c}$}{Speed and mobile direction of mobile scatterer, respectively.}%
\nomenclature{$c\left( t \right)$}{Distance between base station (BS) and the first bounce/scatterer at time $t$.}%
\nomenclature{$a\left( t \right)$}{Distance between the last bounce/scatterer and mobile station (MS) at time $t$.}%
\nomenclature{$\lambda$}{Wavelength.}%
\nomenclature{$\max(\cdot)$}{Maximum.}%
\nomenclature{$\text E\{\cdot\}$}{Statistical expectation operator.}%
\nomenclature{$(\cdot)^*$}{Complex conjugation operation.}%
\nomenclature{$\left\lfloor {\cdot} \right\rfloor$}{Floor function.}%
\printnomenclature

\section{Introduction}
\label{sec:introduction}
\PARstart{T}{he} annual global road crash statistics revealed that road accidents are the direct cause of death for about 1.3 million people every year. Globally, it is estimated to cost USD 518 billion~\cite{WHO 2016}. Therefore, researchers have been focusing on vehicular communication technologies towards accident-free traffic environment. Vehicular communication networks facilitate information sharing between cars and with the surrounding environments. Such information is quite useful for facilitating road safety. During the past few years, considerable research endeavors have been attempted towards the adaptation of vehicle-to-vehicle (V2V), vehicle-to-roadside (V2R), and roadside-to-vehicle (R2V) communications to intelligent transportation systems (ITS). However, ongoing research efforts are focusing on developing the existing technologies toward vehicle-to-everything (V2X) communications. V2X strives towards data sharing between vehicles and homes, pedestrians, grids, devices or other entities that can influence the vehicle.\par

Recently, vehicular ad-hoc networks (VANET) gained enormous attention and become a key part of ITS to reinforce safety on the roads, increase traffic efficiency, and ensure the safety and comfort to drivers, travelers, and passersby~\cite{Zeadally2012}. VANET use dedicated short range communications (DSRC) and wireless access in vehicular environments (WAVE) standards for secure and fast vehicular communications. The implementation of DSCR/WAVE technologies requires new hardware to be added. However, supporting the existing infrastructures and vehicles with new hardware will add extra costs and increase power consumption. On the other hand, visible light communications (VLCs) have attracted ever-growing attention as a complementary technology to radio frequency (RF) based wireless communications for indoor and outdoor wireless environments~\cite{Fourat14}. This brought the idea of exploiting state-of-the-art VLC technique to be integrated with vehicular communications to propose vehicular VLC (VVLC)~\cite{Thesis}. VLC systems take an advantage of commercially available incoherent light-emitting diode (LED) to serve as an optical transmitter (Tx). On the other hand, the optical receiver (Rx) employs a highly sensitive photodiode (PD) or a camera receiver~\cite{A.Al-Kinani}. \par
In order to get optimum VLC system design, explicit knowledge for the optical wireless propagation channel is vital to understand the channel impact on system performance. Considerable research efforts have been carried out related to VVLC channel modeling  in terms of V2R scenarios such as traffic light control at intersections~\cite{Wada 05}\hspace{0.3mm}--\hspace{0.3mm}\cite{Iwasaki2008}. But only the line-of-sight (LoS) channel was taken into account for specific scenarios of applications. These models are deterministic and depend solely on the Tx-Rx distance. Whilst in reality, the received signal consists of LoS and non-LoS (NLoS) components. NLoS components result from reflections off the surrounding obstacles. Furthermore, ray-tracing based channel models for V2V and V2R VVLC channels were proposed in \cite {Lee2012},~\cite {S. Lee2012}. Ray-tracing channel modeling is a deterministic and reliable but time-consuming approach and cannot be extended to a broad range of scenarios~\cite{AhmedS}. In~\cite{Luo2015} the authors considered measurements campaign in~\cite{I-Table}, where both LoS and NLoS links are considered using a geometry-based road-surface reflection channel model. However, the model takes into account the reflections off the blacktop while ignoring other reflections from the surrounding vehicles. As an attempt to fill the above research gap, we previously proposed a two-dimensional (2D) non-stationary regular-shaped geometry-based stochastic model (RS-GBSM) for VVLC single-input single-output (SISO) channels~\cite{2D SISO}. The proposed model considers the LoS and NLoS links and takes into account the surrounding vehicles and the stationary roadside environments. In order to develop the existing 2D RS-GBSM, we propose a three-dimensional (3D) RS-GBSM for the sake of more accurate characterization of VVLC channel models. Due to their reasonable complexity and mathematical traceability, 3D RS-GBSMs have been utilized to investigate channel characteristics of conventional RF-based vehicular channels, as reported in~\cite{Yuan2015}\hspace{-0.01mm}--\hspace{-0.01mm}\cite {Yuan14}. However, to the best of the authors$’$ knowledge, this work presents the first ever efforts to investigate 3D RS-GBSM VVLC channels.

To summarize, the main contributions and novelties of the paper are highlighted as follows:
\begin{enumerate}
\item A 3D RS-GBSM is proposed for VVLC multiple-input single-output (MISO) channels considering the surrounding moving vehicles and stationary roadside environment.
\item We utilize the proposed 3D RS-GBSM to drive and investigate VVLC channels' characteristics such as received optical power and signal-to-noise ratio (SNR).
\item The main differences between conventional RF-based vehicular and VVLC systems are addressed.
\item The proposed 3D RS-GBSM is compared with the existing 2D RS-GBSM.
\item We also investigate the impacts of von-Mises-Fisher (VMF) distribution parameters and elevation angle on VVLC channel characteristics.
\end{enumerate}

The rest of this paper is structured as follows. Section~\ref{Section_II} describes the proposed VVLC MISO system model including the headlamp (Tx) and optical receiver (Rx) models. In Section~\ref{Section III}, the description of the proposed 3D RS-GBSM and the derivations of channel parameters are presented. Section~\ref{Section_IV} presents the investigated VVLC channel characteristics using von Mises-Fisher (VMF) distribution. Numerical and simulation results are shown and analyzed in Section~\ref{Section_V}. Finally, conclusions are made in Section~\ref{Section_VI}.

\section{VVLC MISO System Model}\label{Section_II}
\subsection{VVLC System Model}
Compared with classical RF-based vehicular communication systems, VVLC is classified as small spatial scale (SSS) communication scenario since the Tx-Rx distance is between 30 and 300 m \cite{Wang09}. Unlike conventional RF V2V communication systems, VLC employ intensity modulation~(IM) technique since incoherent LEDs cannot directly be phase or frequency modulated~\cite{Dixit2012}. For signal recovery, direct detection (DD) technique is used. Table~\ref{Table3} presents the key differences between the conventional RF (DSRC) vehicular systems and VVLC systems. \par
New cars have front, tail, and wing mirrors indicator LED lights that can be used as Txs. While a PD or a camera-based receiver can serve as a Rx. Fig.~\ref{Fig1} shows a typical geometrical description of the proposed VVLC scenario with LoS and single-bounced (SB) rays. In this paper, only SB rays are considered since the powers of double-bounced (DB) rays are significantly low and can be disregarded especially for outdoor VLC applications as we have demonstrated in our previous work~\cite{2D SISO}. \par
In general, to model VVLC channels, road traffic and light propagation need to be modeled considering Tx radiation pattern and Rx aperture size~\cite{Maurer2005}. In terms of road traffic modeling, Fig.~\ref{Fig1} illustrates a VVLC system model utilized in urban canyon environments.

\begin{table*}[t]
\normalsize
\centering
\caption{Comparison of VVLCs and RF (DSRC) technologies.}
\label{Table3}
\begin{tabular}{|c|c|c|}
\hline
&\textbf{VVLC} & \textbf{RF (DSRC)}                                                                                                                                                                                                                  \\ \hline
\begin{tabular}[c]{@{}c@{}}Communication\\ Scenario\end{tabular}                              & Mainly (LoS)                              & LoS \& NLoS                                                       \\ \hline
Distance                                                                                                                               & SSS                                        &  SSS, MSS, LSS                                                     \\ \hline
Cost                                                                                                                                     & Low                                           & High                                                                    \\ \hline
\begin{tabular}[c]{@{}c@{}}New Hardware\\ Required \end{tabular}                            &No                                             & Yes                                                                      \\ \hline
Complexity                                                                                                                          & Low                                           & High                                                                    \\ \hline
\begin{tabular}[c]{@{}c@{}}Positioning\\ Precision\end{tabular}                                    & High (cm-level)                          & Low                                                                     \\ \hline
Interference                                                                                                                       & \begin{tabular}[c]{@{}c@{}} Optical (High)\\ Electrical (Low) \end{tabular}                    &Electrical (High)     \\ \hline
\begin{tabular}[c]{@{}c@{}}Environment-\\friendly \end{tabular}                                 & Yes                                          & No                                                                                                                    \\ \hline
Data Rate                                                                                                                           & hundreds of Mbps                             &27~Mbps                                                                           \\ \hline
\begin{tabular}[c]{@{}c@{}} Carrier\\ frequency \end{tabular}                                     & 380--780~THz                             & 5.85--5.925~GHz                            \\ \hline
License                                                                                                                                & Free                                    & Required                                               \\ \hline
Mobility                                                                                                                                & Low-Medium & High         \\ \hline
Security                                                                                                                               & High & Low         \\ \hline
\multicolumn{3}{|l|}{\begin{minipage}[t]{4in}
SSS (Tx-Rx distance $< 300$~m); MSS:~Moderate spatial scale (1~km$>$Tx-Rx distance $>300$~m); LSS:~large spatial scale (Tx-Rx distance $>1$~km)~\cite{Wang09}  \end{minipage}} \\\hline
\end{tabular}
\end{table*}
We assume that there are effective scatterers positioned on 3D ordinary shapes, namely, two spheres and an elliptic-cylinder. The two-sphere model proposes the Tx-sphere and Rx-sphere which are shaping the effective scatterers around the Tx and Rx, respectively. These scatterers represent adjacent moving vehicles. While the elliptic-cylinder model is proposed to model the stationary roadside environments such as architectures, road signages, parked vehicles, and trees. An effective scatterer can involve several closely located physical scatterers that are unresolvable in the delay domain~\cite{Yuan14}. On the other hand, regarding light propagation from the Tx to the Rx, we assume that both the left-side headlight (LSH) and right-side headlight (RSH) have identical output light distribution. Consequently, the received power is composed of LoS and NLoS components. It is worth mentioning that the NLoS components are due to the reflection of LSH and RSH lights off both two-sphere and elliptic-cylinder models. Since there are two headlights at the Tx vehicle and a specific one Rx at the target vehicle, the proposed system is assumed as a MISO system model. \par
\Figure[t!][scale=0.38]{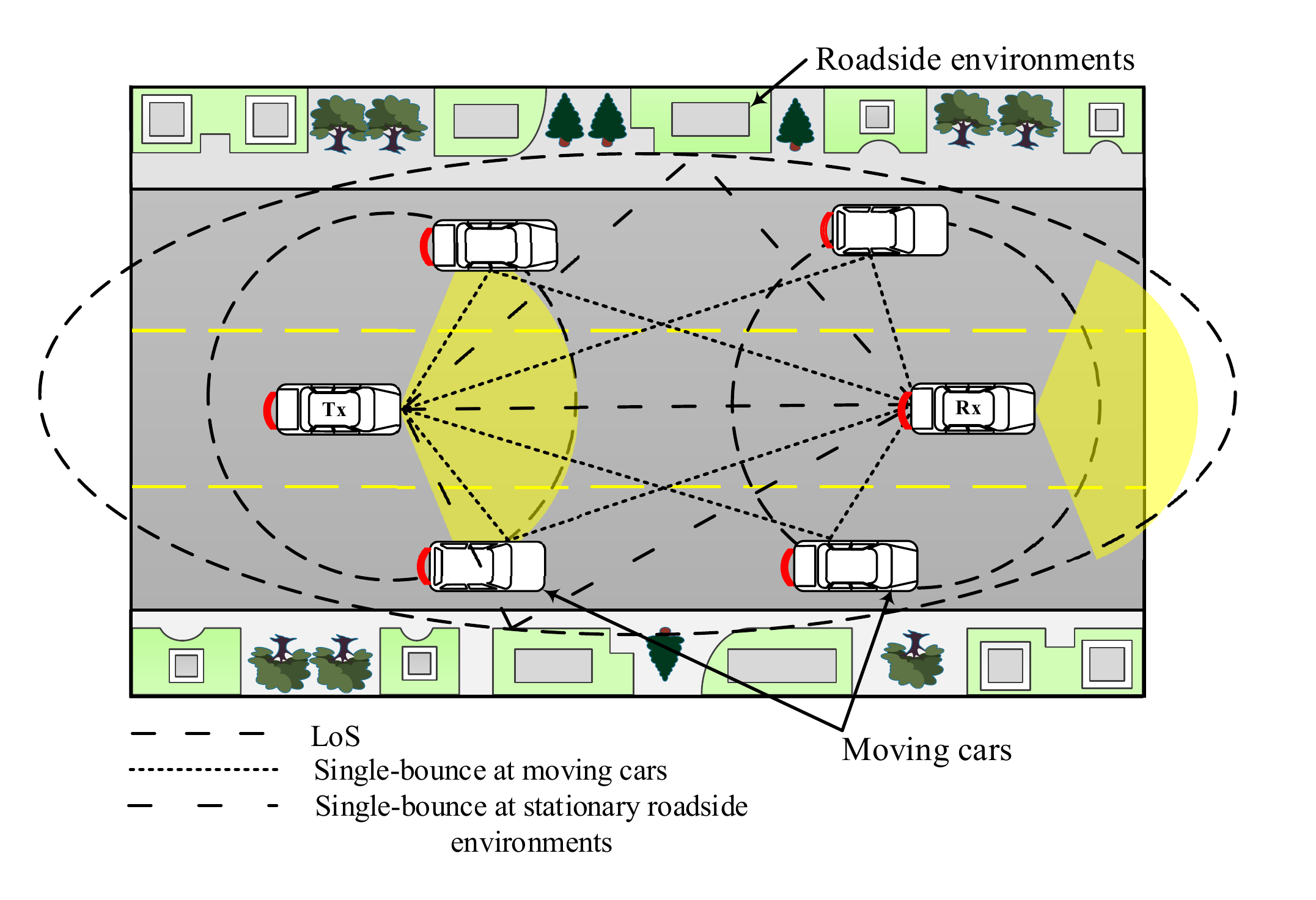}
{The proposed VVLC MISO system model.\label{Fig1}}
In order to introduce the problem of MISO channel modeling, we assume that the VVLC MISO system consists of LSH and RSH headlights with transmit powers of $P_{\mathrm{Tx-\mathrm{LSH}}}$ and $P_{\mathrm{Tx-\mathrm{RSH}}}$, respectively. While at the receiver side, a non-imaging P-type/Intrinsic/N-type (PIN) PD is considered. 
For optical wireless communications (OWC), channel effect is characterized by its impulse response $h(t)$, which is also indicated as channel impulse response (CIR)~\cite{Barry1993}. Optical CIR $h(t)$ expresses the optical power loss and hence plays a significant role to analyze channel effect on VVLC system performance.  With regard to optical channels, the direct current (DC) channel gain $H(0)$ for an optical receiver is given as~\cite{Barry1994}
\begin{equation}
 H(0) =  \int_{0}^{\infty } h(t)dt.
\label{DC_Gain}
\end{equation}
DC channel gain $H(0)$ is used to characterize the losses of the optical channels.\\
Since each Tx will be connected with the Rx through transmission links (sub-channels), therefore, the detailed CIR of the LSH and RSH can be expressed as 
\begin{equation}
\begin{aligned}
h(t)_{\mathrm{L}}=h_{\mathrm{L}}^{\mathrm{LoS}}(t)+\sum_{n=1}^{N} h_{\mathrm{L}}^{i,j}(t)+\sum_{n=1}^{N} h_{\mathrm{L}}^{i,j}(t)+\sum_{n=1}^{N} h_{\mathrm{L}}^{i,j}(t)
\label{h-L}
\end{aligned}
\end{equation}
and
\begin{equation}
\begin{aligned}
h(t)_{\mathrm{R}}=h_{\mathrm{R}}^{\mathrm{LoS}}(t)+\sum_{n=1}^{N} h_{\mathrm{R}}^{i,j}(t)+\sum_{n=1}^{N} h_{\mathrm{R}}^{i,j}(t)+\sum_{n=1}^{N} h_{\mathrm{R}}^{i,j}(t).
\label{h-R}
\end{aligned}
\end{equation}
Due to pages limit, ${\mathrm{L/R}}$ denotes ${\mathrm{LSH/RSH}}$ throughout this paper. Here, $i=1,2$ means we consider the contributions from both left side and right side surroundings for each headlight, namely, $i=1$, for the left side surroundings, while $i=2$ for the right side surroundings. On the other hand, $j=1,2,3$ denotes there are three components for SB  rays, which arrive from the Tx-sphere, the Rx-sphere, and the elliptic-cylinder models, respectively. \par
The total received power for the proposed  MISO VVLC system is generally defined as
\begin{equation}
P_{\mathrm{Rx}}= H(0)_{\mathrm{LSH}}~P_{\mathrm{Tx-\mathrm{LSH}}}+H(0)_{\mathrm{RSH}}~P_{\mathrm{Tx-\mathrm{RSH}}}.
\label{Tot-P1}
\end{equation}
Here, $H(0)_{\mathrm{LSH}}$ and $H(0)_{\mathrm{RSH}}$ represent the DC channel gains of the left headlight and right headlight, respectively.
If we assume that both headlights are transmitting the same power, Eq.(\ref{Tot-P1}) can be rewritten as
\begin{equation}
P_{\mathrm{Rx}}= P_{\mathrm{Tx}}  \big\{H(0)_{\mathrm{LSH}}+H(0)_{\mathrm{RSH}}\big\}.
\label{Tot-P2}
\end{equation}
In this regards, Eq.(\ref{Tot-P2}) represents the most general equation for describing the received optical power of the proposed system model.\par
\subsection{Headlamp Model}
According to the final report from the European Commission, the advanced headlights must be designed to maximize clarity of the roadway whilst minimizing the glare towards oncoming vehicles~\cite{European Commission}. Therefore, the pattern of light produced by a headlamp is of vital importance in VVLC. Headlamps can produce high-beam pattern for long-distance visibility on roads with no oncoming car and low-beam pattern which provides maximum forward and lateral illumination. In this work, we consider low-beam headlight since our system model has been proposed in a typical urban canyon environment. Prior to introducing headlights' radiation pattern, it is important to introduce the unit of measurement of light, which is used to measure lighting level. In terms of surface information, the total luminous flux falling on a unit area of a surface is termed illuminance~($E$) or illumination. Illuminance unit of measurement is lumen per square meter and commonly called lux (lx). The illuminance~$E$ that can be captured at a specific Rx located at a specific distance of interest can be expressed~as~\cite{Lindsey1997}
\begin{equation}
E= \frac{I(\alpha_{\mathrm{T}}, \beta_{\mathrm{T}})~\mathrm{cos}(\beta_\mathrm{R})}{d^{2}}.
\label{Eq.5}
\end{equation}  
Here, $I(\alpha_{\mathrm{T}}, \beta_{\mathrm{T}})$ is the luminous intensity in unit of candela~(cd), $\alpha_{\mathrm{T}}$ and $\beta_{\mathrm{T}}$ are the azimuth and elevation angles of Tx radiation pattern, respectively. Tx-Rx distance denoted as $d$, while $\beta_{\mathrm{R}}$ is the angle between the light-receiving surface normal and the light incident direction. For instance, the illuminance pattern of a headlamp equipped with a Xenon lamp is presented in Fig.~\ref{Xenonlamp}. 
This diagram is called an Isolux diagram where lines indicate illuminance $E$ levels in steps. In this example, the illuminance reaches a maximum as 100~lx at the front of the car and minimum of 1~lx at the outer line. However, such illuminance patterns are asymmetrical, therefore usually the value of $I(\alpha_{\mathrm{T}}, \beta _{\mathrm{T}})$ for a  specific luminaire and specific range of $\alpha_{\mathrm{T}}$ and $\beta _{\mathrm{T}}$ can be provided based on measurements campaign to produce what so-called $I$-table. For instance, $I$-table for standard tungsten-halogen headlamp can be found in~\cite{I-Table}. Since the commercially available Halogen and Xenon lamps cannot be intensity modulated, whereas no $I$-table available for advanced LED headlights, we assume that the radiation patterns of both LSH and RSH are following the generalized Lambertian radiation pattern. However, since Lambertian radiation pattern has uniaxial uniformity, this makes it independent of $\alpha_{\mathrm{T}}$ and hence, it can be written as~\cite{Gfeller79} 
 \begin{equation}
I(\beta_{\mathrm{T}})= \frac{m+1}{2\pi } \mathrm{cos} ^{m}\left (\beta_{\mathrm{T}}\right ) ,~~~~~\beta_{\mathrm{T}} \in \left [ -\pi /2,\pi /2 \right ].
\label{Eq.5}
\end{equation}  
Here, $m$ refers to Lambert's mode number of the optical source, where higher $m$ results in higher light directionality. The irradiance elevation angle is denoted by $\beta_{\mathrm{T}}$. 
\Figure[t!][scale=0.27]{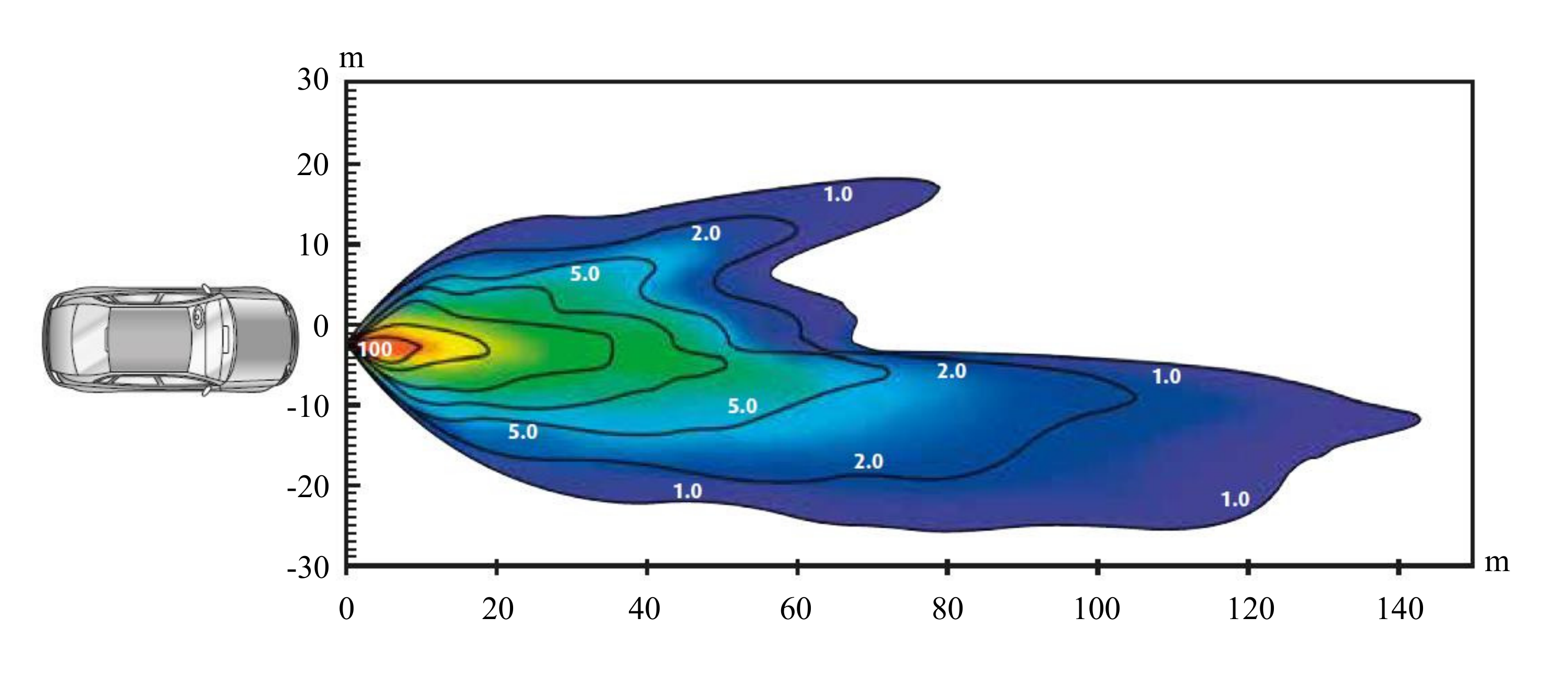}
{An Isolux diagram of a Xenon lamp.\label{Xenonlamp}}
As a validation, simulation results in Fig.~\ref{Lambertianvstungsten} of a Lambertian pattern showed a good match compared to measurements of a tungsten-halogen low-beam headlamp in~\cite{Luo2015}. It should be noted that the Tx and Rx have been mounted at a height~of~0.6~m.
\Figure[t!][scale=0.25]{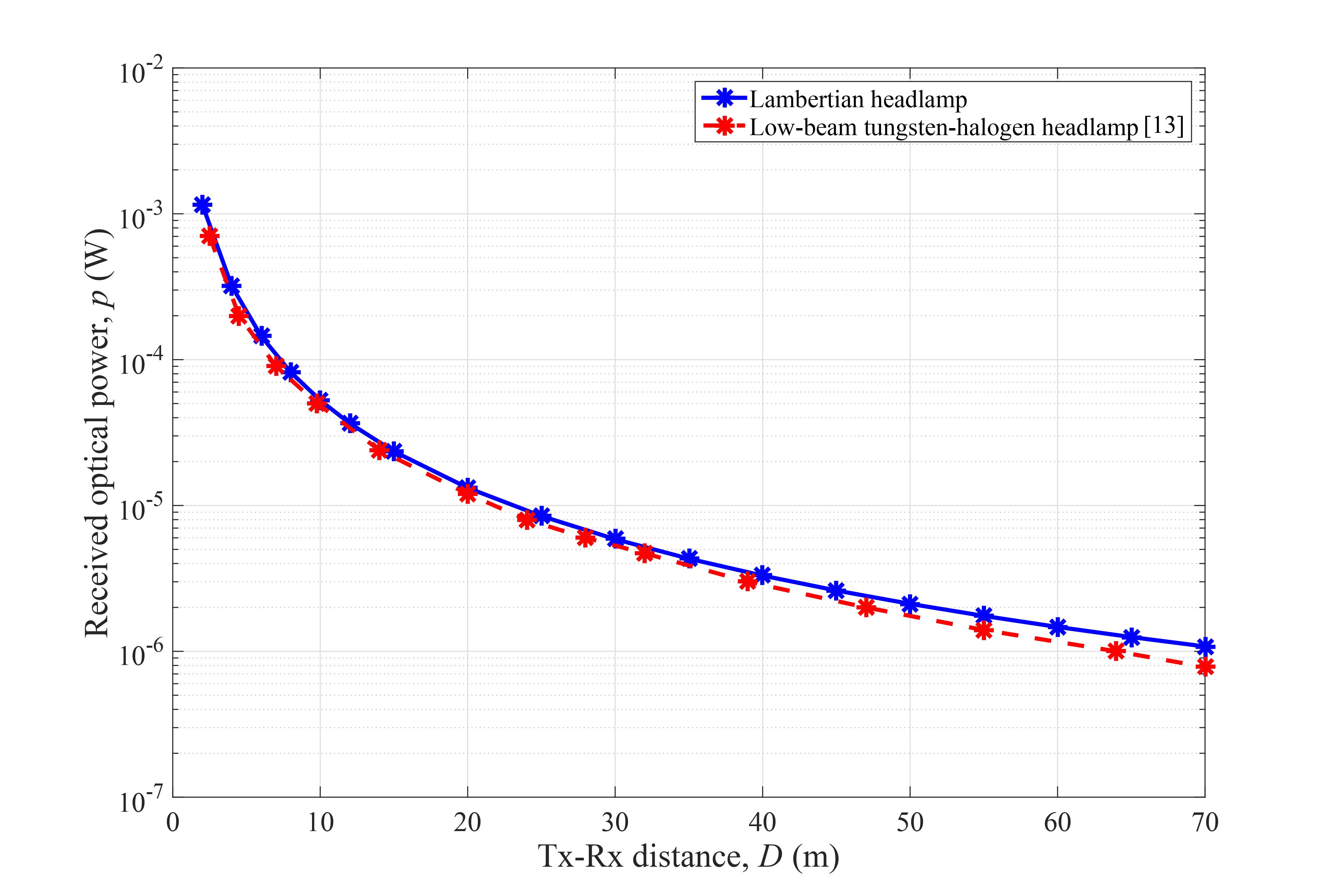} 
{Received power when taking into consideration low-beam tungsten-halogen headlamp in~\cite{I-Table} and Lambertian headlamp, $h_{\mathrm{Rx}}$= 0.6~m.\label{Lambertianvstungsten}}
\subsection{Optical Receiver Model}
The optical receiver can be modeled by its effective area $A_{\mathrm{R}_{_{\mathrm{eff}}}}$, which can be expressed as~\cite{Gfeller79} 
\begin{equation}
A_{\mathrm{R}_{_{\mathrm{eff}}}}=\left\{\begin{matrix}
A_{\mathrm{r}} \cos ( \beta _{\mathrm{R}}),~~~~0\leq\beta _{\mathrm{R}}\leq \mathrm{\Psi _{FoV}} \\0,~~~~~~~~~~~\beta _{\mathrm{R}}>\mathrm{\Psi _{FoV}}.\end{matrix}\right.
\label{Eq.6}
\end{equation}
Here, $A_{\mathrm{r}}$ is the area of the PD. The effective area $A_{\mathrm{R}_{_{\mathrm{eff}}}}$ guarantees that only the light that received within receiver's field of view (FoV) $\mathrm{\Psi _{FoV}}$ will be detected. The effective area can be further extended by attaching a non-imaging concentrator, i.e., a lens to the PD. The optical gain $G(\beta_{\mathrm{R}})$ of the lens is given as~\cite{Ghassemlooy13_Book}
\begin{equation}
G(\beta_{\mathrm{R}})=\left\{\begin{matrix}
\frac{n_{\mathrm{ind}}^2}{\mathrm{sin}^2 (\beta_{\mathrm{R}})},~~~~0\leq \beta_{\mathrm{R}}\leq \mathrm{\Psi _{FoV}} \\0,~~~~~~~ \beta_{\mathrm{R}}> \mathrm{\Psi _{FoV}}.\end{matrix}\right.
\end{equation}
Here, $n_{\mathrm{ind}}$ indicates lens refractive index. An optical filter with $T(\beta_{\mathrm{R}})$ transmission coefficient can be also deposited upon the surface of the lens or integrated to be between the lens and the PD. The optical filter is used to block out-of-band natural and artificial light signals. Using a lens alongside with an optical filter will effectively enhance the detectivity of the PD and reduce undesired ambient light, and hence improve SNR significantly.
\section{The 3D RS-GBSM for MISO VVLC Channels}\label{Section III}
In wireless communications, channel modeling plays a key role as accurate characterization of the propagation channel is essential for a robust communication system design and performance evaluation. In conventional RF-based vehicular communications, RS-GBSMs are widely employed to model V2V channels in 2D and 3D~\cite{Yuan2015}\hspace{-0.01mm}--\hspace{-0.01mm}\cite{Yuan14}, \cite{2D MIMO V2V}\hspace{-0.01mm}--\hspace{-0.01mm}\cite{Wangsurvey}. Therefore, RS-GBSM is applicable even when a different carrier frequency is used. However, utilizing visible light necessitates careful assumptions to adequately capture VVLC channel characteristics. \par
Contrast to wireless RF-based channels, wireless optical-based channels offer high robustness against multipath fading~\cite{Ghassemlooy13_Book}. This is due to the spatial diversity that introduced since the typical PD area is in the order of tens of thousands of optical wavelengths and thus no small-scale fading in OWC. Further, employing of IM/DD technique in OWC systems eliminate frequency offset (FO) between the Tx and Rx since no local oscillators involved. Whereas regarding Doppler shift, is has been reported in~\cite{Elgala2009} that the effect of Doppler frequency in OWC systems is negligible. That is due to a slight corresponding wavelength shift which leads to assume that bandwidth spreading and SNR variation due to Doppler are insignificant problems in most IM/DD systems. In spite of the fact that OWC induce a high robustness against multipath fading, optical channels still experience multipath dispersion, which results in intersymbol interference (ISI).\par
In this study, the wireless optical propagation environment is characterized by a 3D effective scattering with LoS and NLoS components between the Tx and Rx. Fig.~\ref{Ellipsoid} and Fig.~\ref{Spheres} illustrate the proposed 3D non-stationary RS-GBSM for VVLC MISO channels. This model combines the LoS component, two SB components in two-sphere model, and one SB component  in elliptic-cylinder model. For readability purposes, Fig.~\ref{Ellipsoid} only shows the geometry of the LoS and SB components in the elliptic-cylinder model. The geometry of the SB components in the two-sphere model is detailed in Fig.~\ref{Spheres}. 
In order to describe the proposed model, let assume that the Txs are surrounded by a sphere of radius $R_\mathrm{T}$ and there are $ N_1$ effective scatterers are lying on this sphere, where $ n_1$th $(n_1=1,...,N_1)$ is an effective scatterer denoted by $S_{n_{1}}$. Likewise, suppose that the Rx is surrounded by a sphere of radius $R_\mathrm{R}$ and there are $ N_2$ effective scatterers are lying on this sphere, where $ n_2$th $(n_2=1,...,N_2)$ is an effective scatterer denoted by $S_{n_{2}}$. On the other hand, for the elliptic-cylinder model, we assume that  there are $N_3$ effective scatterers are lying on an elliptic-cylinder. Here, the $n_3$th $ (n_3=1,...,N_3)$  local scatterer is denoted by $S_{n_{3}}$. In the latter model, the mid-distance between the Txs, i.e., $\mathrm {OTx}$ and Rx are located at the foci of the elliptic-cylinder. The ellipse parameters $a$ and $b$ (assuming $b<a$) are denoting the semi-major axis and semi-minor axis, respectively. The distance between $\mathrm{OTx}$ (mid-distance between Txs) and the Rx is $D=2f$. Here, $f$ is the half distance between the two focal points of the ellipse and the equality $a^{2}=b^{2}+f^{2}$ holds. Here, the focal points (foci) coincide with firstly, the mid-distance between the two headlights at transmission side and secondly, with the position of the Rx at the receiving side. Table~\ref{Probable optical paths} presents the potential optical ray paths, while the parameters in Figs.~\ref{Ellipsoid} and 5 are defined in Table~\ref{Definitions of key geometry parameters}.
\Figure[t!][scale=0.28]{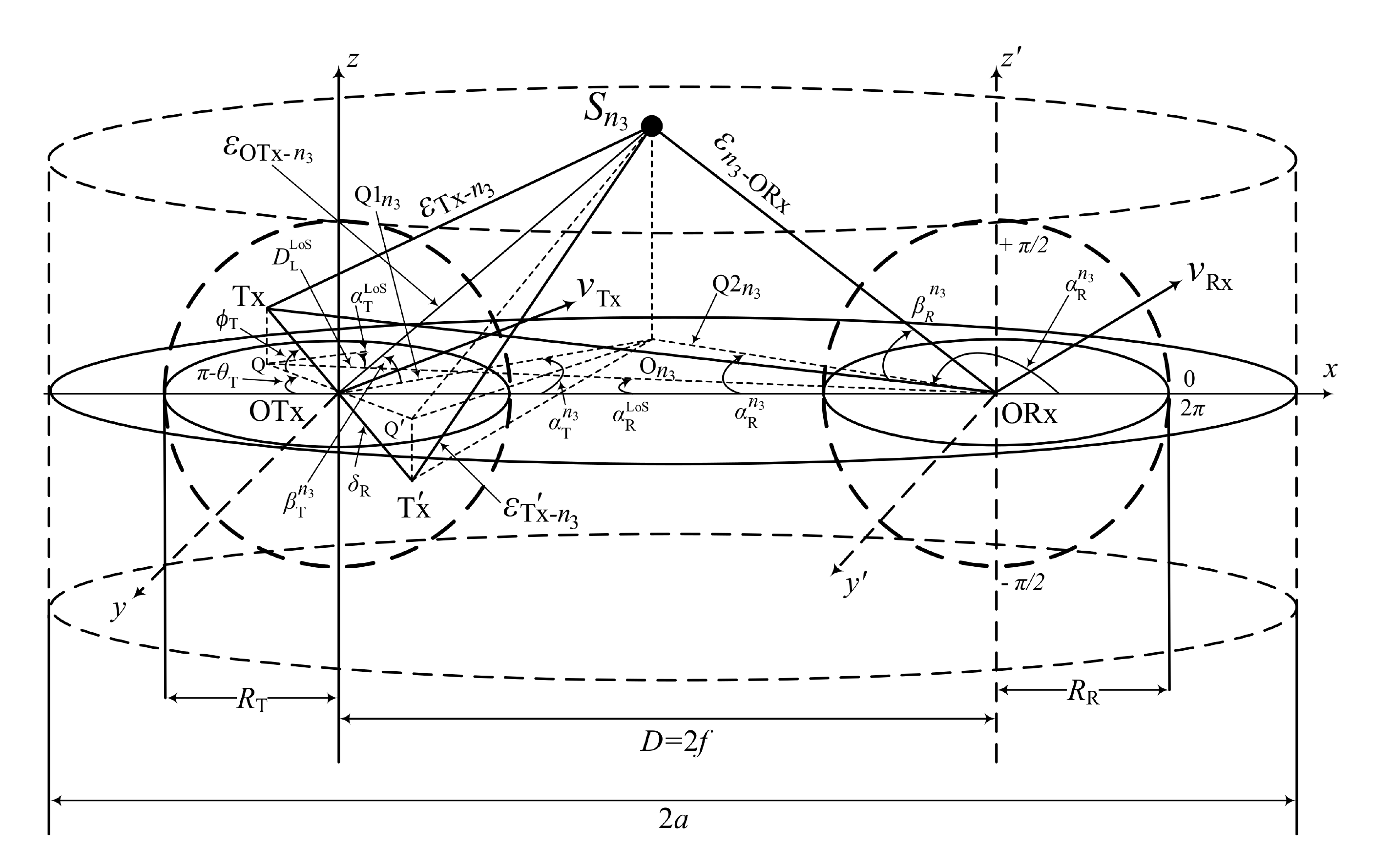} 
{The proposed 3D RS-GBSM for VVLC MISO channels (only the LoS and SB components in elliptic-cylinder model).\label{Ellipsoid}}
\Figure[t!][scale=0.3]{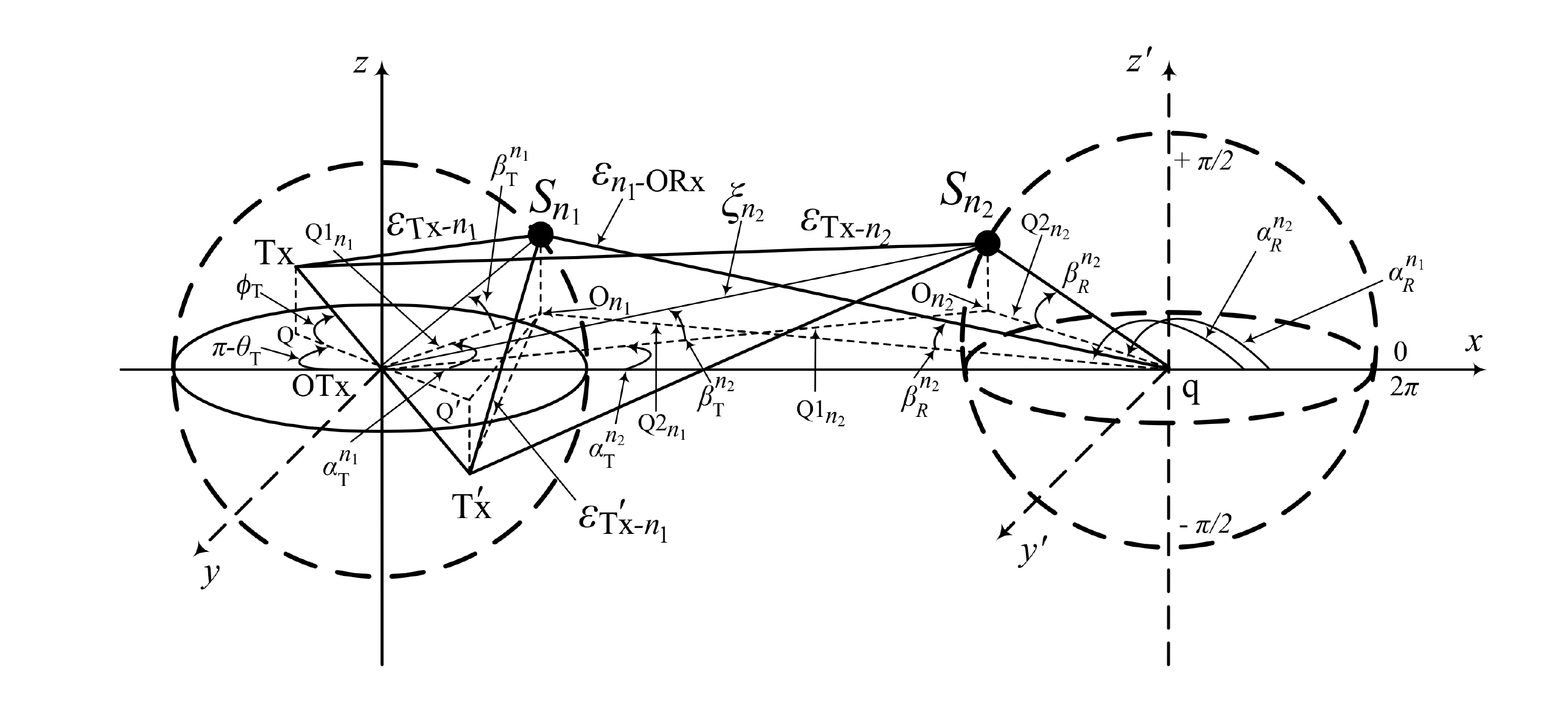} 
{The proposed 3D RS-GBSM for VVLC MISO channels (Only the SB rays in the two-sphere model.\label{Spheres}}
\begin{table}[t]
\caption{The potential optical ray paths.}
\center
\bgroup
\def\arraystretch{1.3}
\begin{tabular}{|c|c|c|}
\hline
\textbf{Component} & \textbf{Optical Path}                                                 & \textbf{Distance} \\ \hline
LoS& $\mathrm{Tx}_{\mathrm {L/R}}$$ \rightarrow $${\mathrm{Rx}}$      &$D$                                                                                      \\ \hline

1- ${\mathrm{SB1}}$ & $\mathrm{Tx}_{\mathrm {L/R}}\rightarrow S_{n_{1}} \rightarrow {\mathrm{Rx}}$ &\begin{tabular}[c]{@{}c@{}}$\varepsilon _{\mathrm{Tx}-n_{1}}+\varepsilon _{n_{1}-\mathrm{ORx}}$,\\ $\varepsilon _{\mathrm{Tx}'-n_{1}}+\varepsilon _{n_{1}-\mathrm{ORx}}$ \end{tabular} \\ \hline

2- ${\mathrm{SB2}}$ & $\mathrm{Tx}_{\mathrm {L/R}}\rightarrow S_{n_{2}} \rightarrow {\mathrm{Rx}}$ &\begin{tabular}[c]{@{}c@{}}$\varepsilon _{\mathrm{Tx}-n_{2}}+R_{\mathrm{R}}$,\\ $\varepsilon _{\mathrm{Tx}'-n_{2}}+R_{\mathrm{R}}$ \end{tabular} \\ \hline

3- ${\mathrm{SB3}}$ & $\mathrm{Tx}_{\mathrm {L/R}}\rightarrow S_{n_{3}} \rightarrow {\mathrm{Rx}}$ &\begin{tabular}[c]{@{}c@{}}$\varepsilon _{\mathrm{Tx}-n_{3}}+\varepsilon _{n_{3}-\mathrm{ORx}}$,\\ $\varepsilon _{\mathrm{Tx}'-n_{3}}+\varepsilon _{n_{3}-\mathrm{ORx}}$ \end{tabular} \\ \hline

\end{tabular}
\egroup
\label{Probable optical paths}
\end{table}
\begin{table*}[t]
\caption{Definition of parameters in Figs.~\ref{Ellipsoid} and 5.} 
\center
    \begin{tabular}{|c|c|}
    \hline
    $D$  & distance between the center of Tx-sphere and the center of the Rx-sphere \\ \hline
    $R_\mathrm {T}, R_\mathrm {R}$ & radius of the Tx and Rx spheres, respectively  \\ \hline
    $ 2\delta $ & spacing between the LSH and the RSH \\ \hline
    $a$, $b$ & semi-principal axes of the ellipse  \\ \hline
    $\theta _\mathrm {T}, \theta _\mathrm {R}$& orientation of the Tx and Rx the x-y plane, respectively \\ \hline
    $ \phi _\mathrm {T}, \phi_\mathrm {R}$ & elevation of the Tx and Rx relative to the x-y plane, respectively \\ \hline
    $v _{\mathrm{T}}, v_{\mathrm{R}}$ & the speeds of the Tx and Rx, respectively \\ \hline
    $\gamma _{\mathrm{T}}, \gamma _{\mathrm{R}}$ & moving directions of the Tx and Rx in the x-y plane, respectively \\ \hline
    $\alpha ^{\mathrm{LoS}}_{\mathrm{T}}$, $\alpha ^{\mathrm{LoS}}_{\mathrm{R}}$     & AAoD and AAoA of the LoS paths, respectively  \\ \hline
    $\alpha _{\mathrm{T}}^{(n_{i})} (i=1,2)$                                                                          & azimuth angle of departure (AAoD) from the Tx to the effective scatterers $s^{(n_{i})}$ \\ \hline
    $\alpha _{\mathrm{R}}^{(n_{i})} (i=1,2)$                                                                                          & azimuth angle of arrival (AAoA) from the effective scatterers $s^{(n_{i})}$ to the Rx      \\ \hline
    $\beta_{\mathrm{T}}^{(n_{i})} (i=1,2)$                                                                                             & EAoD from the Tx to the effective scatterers $s^{(n_{i})}$  \\ \hline
    $\beta_{\mathrm{R}}^{(n_{i})} (i=1,2)$                                                                                            & EAoA from the effective scatterers $s^{(n_{i})}$ to the Rx  \\ \hline

    $\varepsilon _{\mathrm{Tx}-{n_{i}}}$ & distances from the Tx to scatterers ${n_{i}}, (i=1,2,3)$  \\ \hline
    $ \xi_{\mathrm{OTx}-n_{3}}$ & distances from the center of Tx-sphere to scatterer ${n_{3}}$  \\ \hline
    $\xi_{n_{i}-\mathrm{ORx}}$ & distances from scatterer ${n_{i}}, (i=1,2,3)$ to the Rx  \\ \hline
    \end{tabular}
    \label{Definitions of key geometry parameters}
    \end{table*}
To make the proposed 3D RS-GBSM more realistic and practical, two assumptions have been set. Firstly, the bounced rays reflect off the local scatterers from the far to the near relative to the Rx. In other words, the scatterers behind the Tx and prior to the Rx will be neglected. Secondly, ignoring the rays which are out of the PD's FoV~$\mathrm{\Psi _{FoV}}$. Consequently, some bounced components will not be necessarily taken into account. Therefore, the total channel gain can be represented as a superposition of the optical waves coming from the direct direction, i.e., LoS and NLoS directions which are determined by the mean direction of the local scatterers, as detailed in next subsections.\par
\subsection{The LoS Link}
Since we assume that both LSH and RSH have identical output light patterns, i.e., Lambertian pattern, the detailed derivations for the LoS link contribution will be presented here. For the proposed channel model, the CIR will be deterministic if both the Tx and Rx are static. Accordingly, the received power is proportional to the square of the distance between the Tx and Rx (the inverse square law), PD's area~$A_\mathrm {r}$, the LoS elevation angle of departure (EAoD)~$\beta^{\mathrm{LoS}}_{\mathrm{T}}$, and the LoS elevation angle of arrival (EAoA)~$\beta^{\mathrm{LoS}}_{\mathrm{R}}$. Therefore, Eq.$(4.6)$ in \cite{Barry1994} can be expressed as
\begin{equation}
\begin{aligned}
h_{\mathrm{L/R}}^{\mathrm{LoS}}(t)&=\frac{(m+1)~G(\beta_{\mathrm{R}}) T(\beta_{\mathrm{R}}) A_\mathrm {r}}{2\pi (D^\mathrm{LoS}_{\mathrm{L/R}})^2}~\mathrm{cos}^m(\beta _{\mathrm{T,L/R}}^{\mathrm{LoS}})\\
& \times \mathrm{cos}(\beta_{\mathrm{R,L/R}}^{\mathrm{LoS}})~\delta (t-\frac{D^\mathrm{LoS}_{\mathrm{L/R}}}{\mathrm{c}}).
\label{h-static}
\end{aligned}
\end{equation}
Here, $D^{\mathrm{LoS}}_{\mathrm{L/R}}= \sqrt{(\delta _{\mathrm{L/R}})^2+ D^{2}}$, $\delta (.)$ denotes to the Dirac delta function, $T({\phi_\mathrm{R}})$ is the transmission coefficient of an optical band-pass filter, $G({\phi_\mathrm{R}})$ is the gain of the lens, and $\mathrm{c}$ is the speed of light. It should be mentioned that we further assumed that both headlights and the PD are equipped at the same height, namely, 0.6~m. Therefore, Eq.(\ref{h-static}) was written in terms of the LoS EAoD $\beta^{\mathrm{LoS}}_{\mathrm{T}}$, and EAoA $\beta^{\mathrm{LoS}}_{\mathrm{R}}$. 
On the other hand, if the Tx and Rx are moving in the same direction, Eq.(\ref{h-static}) can be rewritten~as
\begin{equation}
\begin{aligned}
h_{\mathrm{L/R}}^{\mathrm{LoS}}(t)&=\frac{(m+1)~G(\beta_{\mathrm{R}}) T(\beta_{\mathrm{R}}) A_\mathrm {r}}{2\pi (D^\mathrm{LoS}_{\mathrm{TR,L/R}}(t))^2}~\mathrm{cos}^m(\beta_{\mathrm{T,L/R}}^{\mathrm{LoS}})\\
& \times \mathrm{cos}(\beta_{\mathrm{R,L/R}}^{\mathrm{LoS}})~\delta (t-\frac{D^\mathrm{LoS}_{\mathrm{TR,L/R}}(t)}{\mathrm{c}}).
\end{aligned}
\label{LoS_CIR_Moving}
\end{equation}
Here, $D^{\mathrm{LoS}}_{\mathrm{TR,L/R}}(t)$ indicates the distance between the Txs and Rx as a function of time. Since the LSH and RSH are located at the same distance from the Rx, $D^{\mathrm{LoS}}_{\mathrm{TR,L/R}}(t)$ can be referred as $D_{\mathrm{TR}}$ and can be given as
\begin{equation}
D^{\mathrm{LoS}}_{\mathrm{TR}}(t)=\varepsilon_{\mathrm{TR}}(t_{0})-[\varepsilon_{\mathrm{Tx}}(t)-\varepsilon_{\mathrm{Rx}}(t)],~~~\upsilon _{\mathrm{Tx}} > \upsilon _{\mathrm{Rx}}.
\label{D-TR-Moving}
\end{equation}
Here, $\varepsilon_{\mathrm{TR}}(t_{0})$, $\varepsilon_{\mathrm{Tx}}(t)$, and $\varepsilon_{\mathrm{Rx}}(t)$ indicate initial distance between the Txs and Rx, the distance of the Txs at the given speed after a specific time, and the target Rx distance at the given speed after a specific time, respectively. If we assume that the motion speed of the Tx and Rx vehicles are $\upsilon _{\mathrm{Tx}}$ and $\upsilon _{\mathrm{Rx}}$, the motion direction will be determined by the angles of motion $\gamma_{\mathrm{Tx}}$ and $\gamma_{\mathrm{Rx}}$, respectively, and hence the distances $\varepsilon_{\mathrm{Tx}}(t)$ and $\varepsilon_{\mathrm{Rx}}(t)$, can be written as $\varepsilon_{\mathrm{Tx}}(t)=\upsilon _\mathrm{Tx}\times t \times \mathrm {cos}(\gamma_\mathrm{Tx})$ and $\varepsilon_{\mathrm{Rx}}(t)=\upsilon _\mathrm{Rx}\times t\times \mathrm {cos}(\gamma_\mathrm{Rx})$, respectively. \par

\subsection{The NLoS Link}
As the number of reflections ${k_{r}}$ increases, determining the CIR  becomes more complex~\cite{Jungnickel2002}. However, the contribution of higher ${k_{r}}$ to the overall outcome is significantly declining since $\left \| h^{k_{r}}(t) \right \|\rightarrow 0,~{k_{r}}\rightarrow \infty$. It has been proven that the primary reflections are dominant over higher order reflections in terms of received power. For instance, indoor measurements have shown that the third bounces carry less than 5\% from the total received power in most scenarios. Consequently, only the primary reflections have been considered in this work. Furthermore, unlike indoor VLC, the outdoor environment is a very different and dynamic and hence more affecting the optical wireless channel characteristics. Accordingly, the power of the second reflection will be quite insignificant as has been demonstrated in~\cite{2D SISO}. Therefore, only the first reflection has been considered in this work. Moreover, to mitigate the complexity of NLoS scenario, the mode number $m$ is assumed to be~1.

\subsubsection{The SB Tx-Sphere Model}
The SB components $h(t)^{(1)}_{\mathrm{L/R}}$ of the CIR within the Tx-sphere model for LSH and RSH can be expressed as 
\begin{equation}
\begin{aligned}
h^{(1)}_{\mathrm{L/R}}(t)&=\lim_{N_{1}\to\infty}\sum_{n_{1}=1}^{N_1} \frac {{G(\beta_{\mathrm{R}}) T(\beta_{\mathrm{R}})A_{\mathrm{r}}~{\rho }_{\mathrm{Vehicles}}}}{\pi^2(\varepsilon _{\mathrm{Tx}-n_{1}})^2~(\varepsilon _{n_{1}-\mathrm{ORx}})^2}\\
& \times \mathrm{cos}(\alpha _{\mathrm{T},{\mathrm{L/R}}}^{(1)})
\mathrm{cos}(\beta _{\mathrm{T},{\mathrm{L/R}}}^{(1)})\\
& \times \mathrm{cos}(\alpha_{\mathrm{R},{\mathrm{L/R}}}^{(1)}) 
\mathrm{cos}(\beta _{\mathrm{R},{\mathrm{L/R}}}^{(1)}) ~\delta (t-\frac{\varepsilon^{(1)}_{\mathrm{L/R}}}{\mathrm{c}}).
\label{h_Tx}
\end{aligned}
\end{equation}

Referring to Fig.~\ref{Spheres}, for the LSH, the distance $\varepsilon^{(1)}_{\mathrm{L}}$ in Eq.(\ref{h_Tx}) can be expressed as
\begin{equation}
\varepsilon^{(1)}_{\mathrm{L}}=\varepsilon _{\mathrm{Tx}-n_{1}}+\varepsilon _{n_{1}-\mathrm{ORx}}.
\label{d(1)}
\end{equation}

The distance $\mathrm{OTx-O_{n_{1}}}$ $(=Q{1}_{n_{1}})$ can be written as $ Q1_{n_{1}}= R_{\mathrm{T}}~\mathrm{cos}(\beta _\mathrm{T,L})$. 
While the distance ${\mathrm{O}_{n_{1}}-\mathrm{ORx}}$ $(=Q{2}_{n_{1}})$ is given as
\begin{equation}
Q{2}_{n_{1}}= \sqrt{(Q{1}_{n_{1}})^2+4f^2-4f(Q{1}_{n_{1}})~{\mathrm{cos}}(\alpha_\mathrm{T,L})}.
\label{d(2)}
\end{equation}
Note that $f$ denotes the distance from the center of the ellipse to each focus. Accordingly, by applying mathematical manipulation, the distance between the LSH and a scatterer, which is lying on the Tx-sphere can be written as
\begin{equation}
\begin{aligned}
\varepsilon _{{\mathrm{T_{x}}}-{n_{1}}}&=(R_{\mathrm{T}}^2+\delta_{L}^2-2 \delta_{L} R_{\mathrm{T}}~\mathrm{cos}(\phi _{\mathrm{T,L}})\\
& \times \mathrm{cos}(\beta  _{\mathrm{T,L}})~\mathrm{cos(\theta _{T,L}-\alpha _{T,L})}\\
&-2 \delta_{L} R_{\mathrm{T}}~\mathrm{sin}(\phi _{\mathrm{T,L}})~\mathrm{sin}(\beta  _{\mathrm{T,L}}))^{0.5}.
\label{Tx-n1}
\end{aligned}
\end{equation}
While the distance between the above scatterer and the ORx can be given as
\begin{equation}
\varepsilon _{{n_{1}}-{\mathrm{OR_{x}}}}=Q2_{n_{1}}/\mathrm{cos(\beta _{\mathrm{R,L}}}). 
\end{equation}
Whilst, with regard to the RSH, the distance between the right headlight and a scatterer, which is lying on the Tx-sphere, can be obtained by applying trigonometry in triangles $\mathrm{OTx}-{\mathrm{Q}}'-\mathrm{O}_{n_{1}}$, ${\mathrm{Q}}'-{\mathrm{Tx}}'-S_{n_{1}}$, and ${\mathrm{Q}}'-{\mathrm{Tx}}'-\mathrm{O}_{n_{1}}$ to get
\begin{equation}
\varepsilon _{{{\mathrm{T_{x}}}}'-{n_{1}}}=\sqrt{R_{\mathrm{T}}^2+\delta _{R}^{2}+A1-B1}
\end{equation}
where, 
\begin{equation}
\begin{aligned}
A1&=2~R_{\mathrm{T}}\delta_{\mathrm{T}}~\mathrm{sin(\phi_{\mathrm{T,R}}})~\mathrm{sin(\beta_{\mathrm{T,R}}})
\end{aligned}
\end{equation}
and 
\begin{equation}
B1=2~R_{\mathrm{T}}\delta_{\mathrm{T}}~\mathrm{cos}(\phi_{\mathrm{T,R}})~\mathrm{cos}(\beta_{\mathrm{T,R}})~\mathrm{cos}(\theta_{\mathrm{T,R}}-\alpha _{\mathrm{T,R}}).
\end{equation}
It is worth mentioning that the azimuth/elevation angle of departure (AAoD/EAoD), (i.e., $\alpha^{(1)}_{\mathrm{T,L/R}}$, $\beta^{(1)}_{\mathrm{T,L/R}}$) and azimuth/elevation angle of arrival (AAoA/EAoA), (i.e., $\alpha^{(1)}_{\mathrm{R,L/R}}$, $\beta^{(1)}_{\mathrm{R,L/R}}$), are correlated for SB rays. Consequently, a relationship between the AoDs and AoAs can be written as
\begin{equation}
\alpha^{(1)}_{\mathrm{R,L/R}}=\arcsin \left ( \frac{R_{\mathrm{T}}~\mathrm{cos}(\beta _\mathrm{T,L/R})~\mathrm{sin}(\alpha _{\mathrm{T,L/R}})}{ \sqrt{(Q{1}_{n_{1}})^2+4f^2+4f(Q{1}_{n_{1}}){\mathrm{cos}}(\alpha_\mathrm{T,L/R})}} \right )
\end{equation}
and
\begin{equation}
\beta^{(1)}_{\mathrm{R,L/R}}=\arctan \left ( \frac{{R_{\mathrm{T}}~\mathrm{sin}(\beta  _{\mathrm{T,L/R}})}}{ \sqrt{(Q{1}_{n_{1}})^2+4f^2+4f(Q{1}_{n_{1}}){\mathrm{cos}}(\alpha_\mathrm{T,L/R})}} \right ).
\end{equation}
\subsubsection{The SB Rx-Sphere Model}
The SB components of the CIR $h(t)^{(2)}_{\mathrm{L/R}}$ within the Rx-sphere model for LSH and RSH, can be written as 
\begin{equation}
\begin{aligned}
h(t)_{\mathrm{L/R}}^{(2)}&=\lim_{N_{2}\to\infty}\sum_{n_{2}=1}^{N_2} \frac {{G(\beta_{\mathrm{R}}) T(\beta_{\mathrm{R}})A_{\mathrm{r}}~{\rho }_{\mathrm{Vehicles}}}}{\pi(\varepsilon^{(2)}_{\mathrm{L/R}})^2}~\mathrm{cos}(\alpha _{\mathrm{T,L/R}}^{{n_{2}}}) \\
& \times \mathrm{cos}(\beta _{\mathrm{T,L/R}}^{{n_{2}}})~\mathrm{cos}(\alpha _{\mathrm{R,L/R}}^{{n_{2}}})~\mathrm{cos}(\beta _{\mathrm{R,L/R}}^{{n_{2}}}) \\
& \times \delta (t-\frac{\varepsilon^{(2)}_{\mathrm{L/R}}}{\mathrm{c}}).
\end{aligned}
\label{Rx-n2}
\end{equation}
For the LSH, the distance $\varepsilon^{(2)}_{\mathrm{L}}$ in (\ref{Rx-n2}) is given~as
\begin{equation}
\varepsilon^{(2)}_{\mathrm{L}}=\varepsilon _{\mathrm{T_{x}}-{{n_{2}}}}+\mathrm{R_{R}}.
\end{equation}
Regarding the optical path lengths within the Rx-sphere model, the distance $\mathrm{OTx-O_{\mathit{n}_{2}}}$ $(=Q1_{n_{2}})$ can be written~as 
\begin{equation}
Q1_{n_{2}}= \sqrt{4f^2+(Q2_{n_{2}})^{2}-4f(Q2_{n_{2}})~\mathrm{cos}(\alpha_R)}.
\end{equation}
 Here, $Q2_{n_{2}}= R_{\mathrm{R}}~\mathrm{cos}(\beta _\mathrm {R})$.
Hence,
\begin{equation}
\xi_{n_{2}}=\sqrt{Q1_{n_{2}} ^2+R_\mathrm{R}^2~\mathrm{sin}^2(\beta _{\mathrm{R}})}.
\end{equation}

The distance between the LSH and a scatterer, which is lying on the Rx-sphere $\varepsilon _{\mathrm{Tx}-n_{2}}$, can be obtained by applying Pythagoras's theorem and the law of sines in the appropriate triangles. Hence, $\varepsilon _{\mathrm{Tx}-n_{2}}$ can be expressed~as 
\begin{equation}
\varepsilon _{\mathrm{Tx}-n_{2}}= \sqrt{A2^{2}+B2^{2}}.
\end{equation}
Here,
\begin{equation}
\begin{aligned}
A2&=(\delta ^{2}~\mathrm{cos}^{2}(\phi _{\mathrm{T}})+(Q{1}_{n_{2}})^{2}\\
&-2 \delta~(Q{1}_{n_{2}})~\mathrm{cos(\phi _{\mathrm{T}}})~\mathrm{cos(\theta  _{\mathrm{T}}-\alpha _{T}}))^{0.5}
\end{aligned}
\end{equation}
and
\begin{equation}
\begin{aligned}
B2&=R_{\mathrm{R}}^{2}~\mathrm{sin}^{2}(\beta _{\mathrm{R}})-2\delta R_{\mathrm{R}}~\mathrm{sin(\beta _{\mathrm{R}}})~\mathrm{sin(\theta  _{\mathrm{T}}})\\
&+\delta ^{2}~\mathrm{sin}^{2}(\phi _{\mathrm{T}}).
\end{aligned}
\end{equation}

With regard to the RSH, the distance between the RSH and a scatterer that lying on the Rx-sphere $\varepsilon _{{{\mathrm{T_{x}}}}'-{n_{2}}}$, can be written~as
\begin{equation}
\begin{aligned}
\varepsilon _{{{\mathrm{T_{x}}}}'-{n_{2}}}=\sqrt{R_{\mathrm{R}}^2~\mathrm{sin}^{2}(\beta _{\mathrm{R}})+\delta ^{2}~\mathrm{sin}^{2}(\phi _{\mathrm{T}})+A3+B3}.
\end{aligned}
\end{equation}
Here,
\begin{equation}
\begin{aligned}
A3&=\delta ^{2}~\mathrm{cos}^{2}(\phi _{\mathrm{T}})+Q1_{n_{2}}^{2}\\
&+2\delta~Q1_{n_{2}}~\mathrm{cos}(\phi _{\mathrm{T}})~\mathrm{cos(\theta  _{\mathrm{T}}-\alpha _{T}})
\end{aligned}
\end{equation}
and
\begin{equation}
B3=2\delta R_{\mathrm{R}} ~\mathrm{sin}(\phi _{\mathrm{T}})\mathrm{cos}(\beta _{\mathrm{R}}).
\end{equation}
Since AAoD/EAoD and AAoA/EAoA are correlated for SB rays in Rx-sphere model, the correlation between the AoDs and AoAs is given by
\begin{equation}
\beta^{(2)}_{\mathrm{T,L/R}}=\arcsin \left ( \frac{R_{\mathrm{R}}~\mathrm{sin}(\beta _\mathrm{R,L/R})}{ \sqrt{(R_{\mathrm{R}}^2+4f^2+4fR_{\mathrm{R}}C3}} \right )
\end{equation}
where,
\begin{equation}
C3=\mathrm{cos}(\beta _\mathrm{R,L/R})~\mathrm{cos}(\alpha_\mathrm{R,L/R})
\end{equation}
and
\begin{equation}
\alpha^{(2)} _{\mathrm{T,L/R}}=\arcsin \left ( \frac{R_{\mathrm{R}}~\mathrm{cos}(\beta _\mathrm{R,L/R})~\mathrm{sin}(\alpha  _\mathrm{R,L/R})}{ Q1_{n_{2}}} \right ).
\end{equation}
\subsubsection{The SB Elliptic-Cylinder Model}
The SB components $h(t)^{(3)}_{\mathrm{L/R}}$ of the CIR within the elliptic-cylinder model for the LSH and RSH can be expressed~as
\begin{equation}
\begin{aligned} 
h_{\mathrm{L/R}}^{(3)}(t)&=\lim_{N_{3}\to\infty}\sum_{n_{3}=1}^{N_3} \frac {{G(\beta_{\mathrm{R}}) T(\beta_{\mathrm{R}})A_{\mathrm{r}}~{\rho }_{\mathrm{Roadside}}}}{\pi(\varepsilon^{(3)}_{\mathrm{L/R}})^2}~\mathrm{cos}(\alpha _{\mathrm{T,L/R}}^{{n_{3}}})\\
& \times \mathrm{cos}(\beta _{\mathrm{T,L/R}}^{{n_{3}}})~\mathrm{cos}(\alpha _{\mathrm{R,L/R}}^{{n_{3}}})~\mathrm{cos}(\beta _{\mathrm{R,L/R}}^{{n_{3}}})\\
& \times\delta (t-\frac{\varepsilon^{(3)}_{\mathrm{L/R}}}{\mathrm{c}}).
\end{aligned}
\label{Elliptic-Cylinder-n2}
\end{equation}
For the LSH in Fig.~\ref{Ellipsoid}, the distance $\varepsilon^{(3)}_{\mathrm{L}}$ in Eq.(\ref{Elliptic-Cylinder-n2}) can be written~as
\begin{equation}
\varepsilon^{(3)}_{\mathrm{L}}=\varepsilon_{\mathrm{T_{x}}-{{n_{3}}}}+\varepsilon _{{n_{3}}-{\mathrm{OR_{x}}}}.
\end{equation}

Within the elliptic-cylinder model, the optical path lengths can be determined by using pure elliptic-cylinder properties that mentioned in above. The distance $\mathrm{OTx-O_{\mathit{n}_{3}}}$ $(=Q{1}_{n_{3}})$ can be expressed as
\begin{equation}
Q1_{n_{3}}= \sqrt{(Q{2}_{n_{3}})^2 +(D)^2-2(Q{2}_{n_{3}})(D)~\mathrm{cos} (\alpha_{\mathrm{R,L}}^{{n_{3}}})}.
\end{equation}
While the distance $\mathrm{O_{\mathit{n}_{3}}-ORx}~(=Q{2}_{n_{3}})$ is given as 
\begin{equation}
 Q2_{n_{3}}=\xi_{n_{3}-\mathrm{ORx}}~\mathrm{cos} (\beta_{\mathrm{R,L}}^{{n_{3}}}).
\label{projector}
\end{equation}

Based on elliptic-cylinder properties and after some manipulation we can get

\begin{equation}
\varepsilon_{n_{3}-\mathrm{ORx}}=\frac{2a-Q_{n_{3}}}{\mathrm{cos}(\beta _{\mathrm{R,L}}^{n_{3}})}
\end{equation}
and 
\begin{equation}
\varepsilon_{\mathrm{OTx}-n_{3}} =\sqrt{Q_{n_{3}}^2+(\xi_{n_{3}-\mathrm{ROx}})^2~\mathrm{sin}^2(\beta _{\mathrm{R,L}}^{n_{3}})}
\end{equation}
where
\begin{equation}
Q_{n_{3}}=\frac{a^2+f^2+2af~{\mathrm{cos}(\alpha _{\mathrm{R,L}}^{n_{3}})}}{a+f~\mathrm{cos}(\alpha _{\mathrm{R,L}}^{n_{3}})}.
\end{equation}
The distance between the LSH and a scatterer that lying on the elliptic-cylinder model $\varepsilon _{{{\mathrm{T_{x}}}}-{n_{3}}}$, can be written as 
 
\begin{equation}
\varepsilon _{\mathrm{Tx}-n_{3}}= \sqrt{A4^2+B4^2}.
\end{equation}
Here, 
\begin{equation}
\begin{aligned}
A4=\delta ^{2}+Q{1}_{n_{3}}^{2}-2\delta~Q{1}_{n_{3}}~\mathrm{cos(\phi _{\mathrm{T}}})~\mathrm{cos(\theta _{\mathrm{T}}-\alpha _{T}})
\end{aligned}
\end{equation}
and
\begin{equation}
\begin{aligned}
B4&=\delta ^2+ \varepsilon_{n_{3}-\mathrm{ORx}}^2~\mathrm{sin}^{2}(\beta _{\mathrm{R}})-2 \delta~\varepsilon_{n_{3}-\mathrm{ORx}}\\
& \times \mathrm{sin}(\beta _{\mathrm{R}})~\mathrm{sin}(\phi _{\mathrm{T}}).
\end{aligned}
\end{equation}

For the RSH, the distance between the RSH and a scatterer that lying on the elliptic-cylinder model $\varepsilon _{{{\mathrm{T_{x}}}}'-{n_{3}}}$, can be written as 

\begin{equation}
\varepsilon _{{{\mathrm{T_{x}}}}'-{n_{3}}}=\sqrt{\delta^{2}~\mathrm{sin^{2}(\phi  _{\mathrm{T}})}+A5-B5}.
\end{equation}
Here,
\begin{equation}
\begin{aligned}
A5=D^2+\varepsilon_{{n_{3}}-\mathrm{ORx}}^2-2D~\varepsilon_{{n_{3}}-\mathrm{ORx}}~\mathrm{cos}(\beta_{\mathrm{R}})~\mathrm{cos}(\alpha _{\mathrm{R}})
\end{aligned}
\end{equation}
 and 
\begin{equation}
B5=2D\delta\varepsilon_{{n_{3}}-\mathrm{ORx}}~\mathrm{sin}(\phi _{\mathrm{T}})~\mathrm{cos}(\alpha_{\mathrm{R}}).
\end{equation}
Since the correlation between AAoD/EAoD and AAoA/EAoA is still valid in elliptic-cylinder model, the relationship between the AAoAs and AAoDs can be expressed~as 
\begin{equation}
\beta_{\mathrm{T,L/R}}^{(3)}=\arcsin \left (  \frac{\varepsilon_{n_{3}-\mathrm{ORx}}~{\mathrm{sin}}(\beta_\mathrm{R,L/R})}{\varepsilon_{\mathrm{OTx}-n_{3}}}\right).
\end{equation}
While the relationship between the EAoAs and EAoDs can be written as
\begin{equation}
\alpha_{\mathrm{T,L/R}}^{(3)}=\arcsin \left (  \frac{\varepsilon_{n_{3}-\mathrm{ORx}}~{\mathrm{cos}}(\beta_\mathrm{R,L/R})~{\mathrm{sin}}(\alpha_\mathrm{R,L/R})}{Q{1}_{n_{3}}}\right).
\end{equation}
\section{VVLC Channel Characteristics}\label{Section_IV}
\subsection{Von Mises Fisher Distribution (VMF)}
Theoretical RS-GBSMs assume infinite number of effective scatterers and hence infinite complexity. However, in this study, only the discrete AAoD $\alpha _\mathrm {T}^{(n_{i})}$, EAoD $\beta _\mathrm {T}^{(n_{i})}$, AAoA $\alpha _\mathrm {R}^{(n_{i})}$, and EAoA $\beta _\mathrm{R}^{(n_{i})}$ will be considered. The methodology of obtaining the set of $\left \{\alpha^{(n_{i})},\beta^{(n_{i})} \right \}$ will be given in the next section. In order to consider the joint impact of the azimuth and elevation angles on channel properties, VMF probability density functions (PDF) is used in this paper to represent the concentration of the effective scatterers. VMF distribution is commonly used to describe directional data and parameterized by a mean direction and a concentration factor $k$. VMF PDF is defined as \cite{Mardia2000}
\begin{equation}
f(\alpha ,\beta )=\frac{k \cos \beta }{4\pi\sinh~k}\mathrm{e}^{k[\mathrm{cos}~\beta _{0}~\mathrm{cos}~\beta~\mathrm{cos}(\alpha -\alpha _{0})+\mathrm{sin}~\beta _{0}~\mathrm{sin}~\beta]}.
\end{equation}
Here, $\alpha\in (-\pi, \pi)$~and~$\beta \in (-\pi/2, \pi/2)$, while $\alpha_{0} \in (-\pi, \pi)$ and $\beta_{0}\in (-\pi/2, \pi/2)$ refer to the mean values of the azimuth angle  $\alpha$  and elevation angle $\beta$, respectively. The $k~(k \geqslant 0)$ parameter is a real-valued parameter which characterizes the concentration of the local scatterers relative to the mean direction, i.e., $\alpha_{0}$ and $\beta_{0}$. In order to demonstrate the VMF on the unit sphere in 3D space, we set $\alpha_{0}=10^{\circ}$, $\beta_{0}=2^{\circ}$, and $k=30$ as an example and plot the scatterers (10000 points) that embedded in a 3D Euclidean space to obtain the distribution that shown in Fig. \ref{3D-2D VMF}(a). While Fig. \ref{3D-2D VMF}(b) illustrates the corresponding VMF PDF. According to VMF distribution, higher values of $k$ imply higher concentration around the direction of the mean angles~\cite{Bian2019}. In consequence, $k\rightarrow 0$ produces an isotropic distribution, while for $k\rightarrow \infty$ the distribution will be extremely non-isotropic.
\subsection{Channel DC Gain}
Let us consider LSH and RSH with Lambertian sources, a receiver with an optical band-pass filter of transmission $T({\phi_\mathrm{R}})$ and a lens of gain $G({\phi_\mathrm{R}})$, the channel DC gain for the LoS links can be expressed~as
\begin{equation}
\begin{aligned}
H(0)_{\mathrm{L/R}}^{\mathrm{LoS}} =\begin{cases}\frac{G(\beta_{\mathrm{R}}) T(\beta_{\mathrm{R}})A_\mathrm {r}}{\pi (D^\mathrm{LoS}_{\mathrm{TR,L/R}})^2} \\\times \mathrm{cos}(\beta_{\mathrm{T,L/R}}^{\mathrm{LoS}}) 
\mathrm{cos}(\beta  _{\mathrm{R,L/R}}^{\mathrm{LoS}})~ &0< \beta  _{\mathrm{R,L/R}} \leqslant  \mathrm{\Psi _{FoV}}\\0 & \beta  _{\mathrm{R,L/R}} >\mathrm{\Psi _{FoV}}.\end{cases}
\end{aligned}
\end{equation}
On the other hand, in order to consider the joint effect of azimuth and elevation angles on the optical wireless channel for the NLoS scenario, we need to consider the gain of all reflected paths by performing the double integral of the 3D VMF PDF, i.e., the volume of Fig.~\ref{3D-2D VMF}(b). Therefore, channel DC gain of LSH and RSH for the NLoS scenario can be written~as
\begin{equation}
\begin{aligned}
H(0)_{\mathrm{L/R}}^{\mathrm{SB}}=\begin{cases}\int_{-\pi/2}^{\pi/2}  \int_{-\pi/2}^{\pi/2}I_{L,R}(\alpha,\beta )\\
\times h_{L,R}(t)\\
 \times~f_{L,R}(\alpha ,\beta)~d\alpha~d\beta & 0< \beta  _{\mathrm{R,L/R}} \leqslant  \mathrm{\Psi _{FoV}}\\0 & \beta  _{\mathrm{R,L/R}} >\mathrm{\Psi _{FoV}}.\end{cases}
\label{DC Gain}
\end{aligned}
\end{equation}
Here, $f_{L,R}(\alpha ,\beta)$ denotes VMF PDF.  It is worth to note that here we apply the second criteria in Section~\ref{Section III} to exclude the rays which are out of the PD's FoV.
\Figure[t!][scale=0.18]{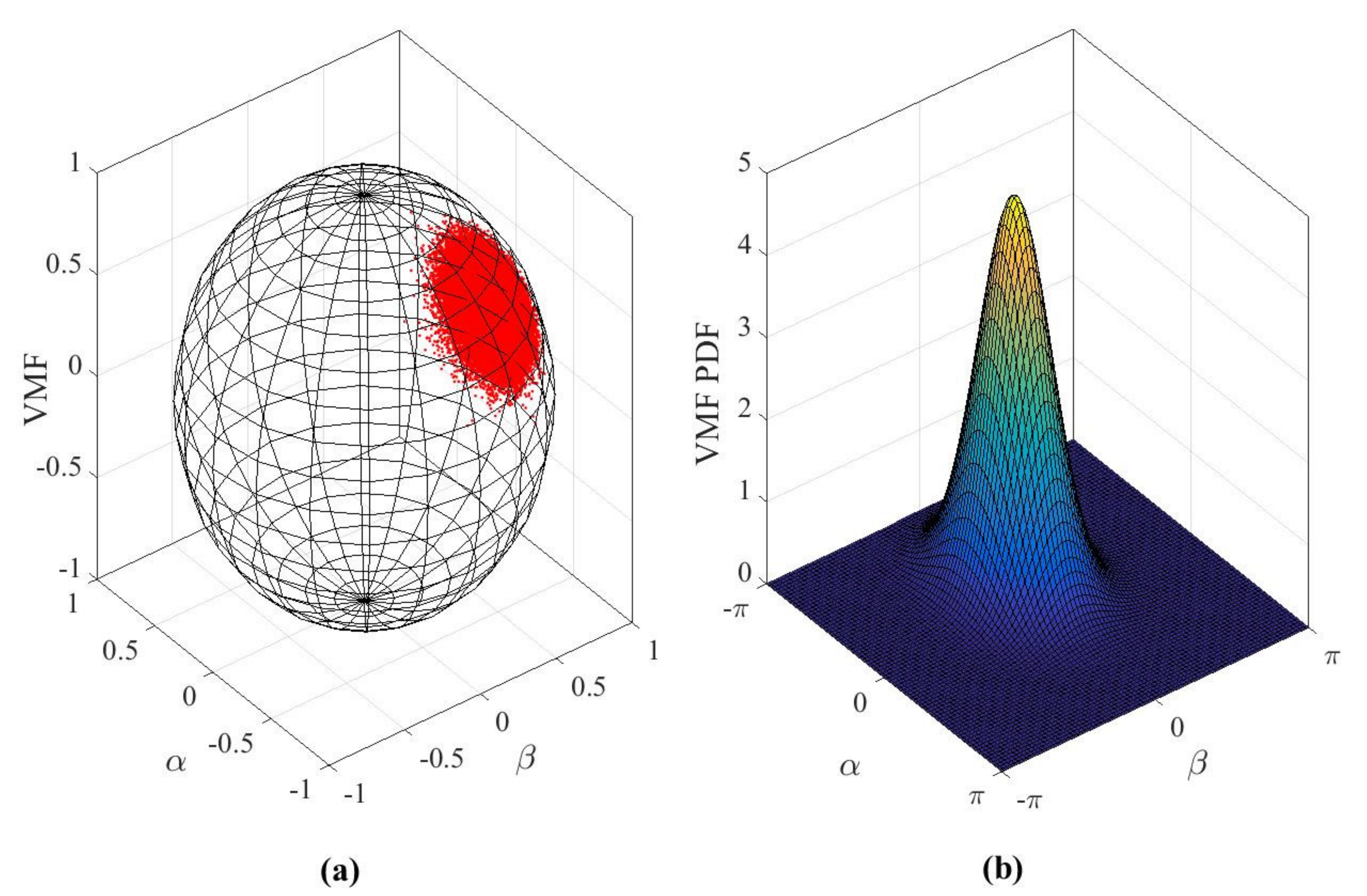} 
{(a) The VMF distribution on the unit sphere in 3D and (b) VMF PDF ($\alpha_{0}=10^{\circ}$, $\beta_{0}=2^{\circ}$, $k = 30$).\label{3D-2D VMF}}
\subsection{Noise Modeling}
For outdoor VLC applications, the optical noise can be produced by background light from solar light during the daytime and other artificial lights such as streetlights, vehicles lights, and advertising screens at nighttime~\cite{Luo2015}. Optical noise is a decisive factor in determining link performance. The total noise at the Rx side is comprised of firstly, the noise that induced by the photocurrent which is known as shot noise~${\sigma_{\mathrm{sh}} ^2}$. Secondly, the noise that resulting from the ambient light sources, i.e., background noise~${\sigma_{\mathrm{b}} ^2}$. The third type of noise is dark current noise ${\sigma_{\mathrm{d}} ^2}$, which is the reverse leakage current induced by a random generation of electrons and holes through the PD in the absence of light. Forthly, the thermal noise ${\sigma_{\mathrm{th}} ^2}$, which is induced by the receiver's electronics such as the resistive elements~\cite{Ahmed dual}. Consequently, the total noise variance is defined as~\cite {Ghassemlooy13_Book}
\begin{equation}
{\sigma_{\mathrm{total}} ^2}={\sigma_{\mathrm{sh}} ^2}+{\sigma_{\mathrm{b}} ^2}+{\sigma_{\mathrm{d}} ^2}+ {\sigma_{\mathrm{th}} ^2}.
\label{total noise}
\end{equation}
The shot noise and thermal noise variances are expressed as \cite{Ghassemlooy13_Book}
\begin{equation}
{\sigma_{\mathrm{sh}} ^2}=2qR_{\lambda } P_{\mathrm{Rx}}B+2qI_{B}I_{2}B
\label{Shot}
\end{equation}
and
\begin{equation}
{\sigma_{\mathrm{th}} ^2}=\frac{8\pi k_{\mathrm{B}}T_{\mathrm{k}}}{G_{\mathrm{ol}}} C_{\mathrm{PD}}A_{\mathrm{r}}I_{2}B^{2}+ \frac{16\pi^{2} k_{\mathrm{B}}T_{\mathrm{k}}\Gamma}{g_{\mathrm{m}}} C_{\mathrm{PD}}A_{\mathrm{r}}^{2}I_{3}B^{3},
\label{Thermal}
\end{equation}
respectively. The other noise contributions in Eq.(\ref{total noise}) can be obtained according to~\cite {Ghassemlooy13_Book} (Eq.(4.7)). In this paper, we have adopted IM/DD that employing on-off keying (OOK) scheme. Therefore, the SNR at the receiver side is given as~\cite{Ghassemlooy13_Book}
\begin{equation}
\mathrm{SNR}=\frac{( R_{\lambda}~P_{\mathrm{Rx}})^2}{\sigma_{\mathrm{total}} ^2}.
\label{SNR}
\end{equation}
\section{Simulation Results and Analysis}\label{Section_V}
In performing simulations, the key parameters for the proposed system model are summarized in Table \ref{Model Parameters_3D}. The most cars have bodies made from either steel or aluminum. For painted steel bodies, the average reflectance~${\rho }_{\mathrm{Vehicles}}$ will be taken into account. On the other hand, for the roadside environment, average concrete reflectance~${\rho }_{\mathrm{Roadside}}$ has been selected. The most important VVLC channel characteristics have been studied in below subsections.
\subsection{Received Optical Power}
In this section, the received wireless optical power is analyzed based the proposed VVLC MISO channel model parameters.
\subsubsection{LoS components}
In this model we consider that the Tx and Rx are moving in the same direction. Since the drivers try to keep the car centered in the current lane, we assume that the reference projection of the Tx vehicle is the lane's center as shown in Fig.~\ref{Fig1}. The target vehicle Rx can be located either in the same lane or in an adjacent lane. The received power will be determined mainly by the Tx-Rx distance. For simulation purposes, we have set the following values for the main model parameters. The initial distance (at $t=0$) between the Tx vehicle and Rx vehicle is 70 m and they are moving with speed of 6 m/s (21.6 km/h) and 4 m/s (14.4 km/h), respectively, in the same direction, i.e., $\gamma_{\mathrm{Tx}}=\gamma_{\mathrm{Rx}}=0$. Here, the headlight separation $2\delta$ is 1.20~m \cite{Luo2015}. As the Tx vehicle is moving at higher speed than the Rx vehicle, we assumed that the stopping distance (SD) is 6~m~\cite{stoppingdistance}. By considering above parameters and applying Eq.(\ref{LoS_CIR_Moving}) and Eq.(\ref{D-TR-Moving}), the simulation results are shown in Fig.~\ref{Fig_7}. 
\begin{table}
\normalsize
\caption{Values of model key parameters used in the simulations.}
\centering
\label{Model Parameters_3D}
\begin{tabular}{|l|l|}
\hline
\multicolumn{2}{|l|} {\textbf{Model Key Parameters}}                                                                                                                      \\ \hline
Initial  Tx-Rx distance                                                           &  70~m                                                                                   \\ \hline
Semi-major $a$ \& semi-minor $b$ axes                                     &   40 m, 19 m                                                                         \\ \hline
Tx speed $\upsilon _{\mathrm{Tx}}$                                       &  21.6 km/h                                                                                \\ \hline
Rx speed $\upsilon _{\mathrm{Rx}}$                                       & 14.4 km/h                                                                                 \\ \hline
Sphere Radius ($R_\mathrm {T}, R_\mathrm {R}$)                  & 4~m                                                                                           \\ \hline
Lane width                                                                                  & 3.5~m~\cite{Lanewidth}                                                         \\ \hline
Roadside width                                                                           & 2.2~m~\cite{Roadsidewidth}                                                   \\ \hline
Stopping distance (SD)                                                               & 6~m~\cite{stoppingdistance}                                                                                       \\ \hline
Vehicles  Reflectivity (${\rho }_{\mathrm{Vehicles}})$              &  0.8~\cite{reflectivity}                                                                                           \\ \hline
Roadside Reflectivity (${\rho }_{\mathrm{Roadside}})$          &  0.4~\cite{concrete}                                                                                                    \\ \hline 
PD Area                                                                                      &  $1~ \mathrm{cm^2 }  $                                                         \\ \hline
Refractive index ($n_{\mathrm{ind}}$)                                    &  1.5                                                                                          \\ \hline
Optical filter gain ($T(\beta_{\mathrm{R}})$)                          &  1                                                                                  \\ \hline
Luminous intensity ($I$)                                                            & 8830~cd~\cite{I-Table}                                                                                           \\ \hline
PD Field of view $(\mathrm{FoV})$                                           & $80^{\circ}$                                                                            \\ \hline
Number of Scatterers                                                                 & 100                                                                                             \\ \hline
\begin{tabular}[c]{@{}c@{}}Capacitance of PD per \\ unit area~($C_\mathrm {PD}$)\end{tabular}             &  112~pF/cm$^{2}$~\cite{Theory2017}                                                                           \\ \hline
\begin{tabular}[c]{@{}c@{}}  Noise bandwidth factors\\ $I_{2}$ and $I_{3}$ \end{tabular}                        & \begin{tabular}[c]{@{}c@{}}0.562 and \\ 0.0868~\cite{Luo2015}\end{tabular}                                                                    \\ \hline  
FET channel noise factor~($\Gamma$)                                      &          1.5~\cite{Ghassemlooy13_Book}                                                                                             \\ \hline  
Boltzmann's constant~($ k_{\mathrm{B}}$)                              &         $1.38 \times 10^{-23}$~J/K                                                     \\ \hline  
Absolute temperature~$(T_{\mathrm {k}})$                           & 298~K                                                                                                 \\ \hline  
Electric charge~$(q)$                                                                 &    $1.6 \times 10^{-19}$~C                                                          \\ \hline  
Open-loop voltage gain~$(G_{\mathrm{ol}})$                          & 10~\cite{Luo2015}                                                                                                   \\ \hline  
FET transconductance~$(g_{\mathrm{m}})$                             &   30~mS~\cite{Luo2014}                                                               \\ \hline  
VLC system bandwidth~$(B)$                                                         &  20~MHz                                                                                     \\ \hline  
Background noise current~$(I_{\mathrm{B}})$                          &  5100 $\mu$A~\cite{Theory2017}                                                 \\ \hline  
\end{tabular}
\end{table}
This figure illustrates the contribution of each headlight in addition to the total received power, which is the sum of the LSH and RSH powers. It can be seen that the received power depends on the Tx-Rx distance and this behavior becomes more pronounced when the Tx and Rx get closer to each~other. \par
On the other hand, we considered the generalized Lambertian radiation pattern because there is no available measured beam pattern for a standard LED headlamp. Therefore, we examine the effect of mode number~$m$ of Lambertian radiation pattern on the received optical power. By taking into account the total received power from both LSH and RSH, it has been demonstrated that the received power increases as the mode number increases as illustrated in Fig.~\ref{Fig_8}. This is due to the fact that higher mode number provides higher directionality of the optical source and hence more power will be delivered.

\Figure[t!][scale=0.29]{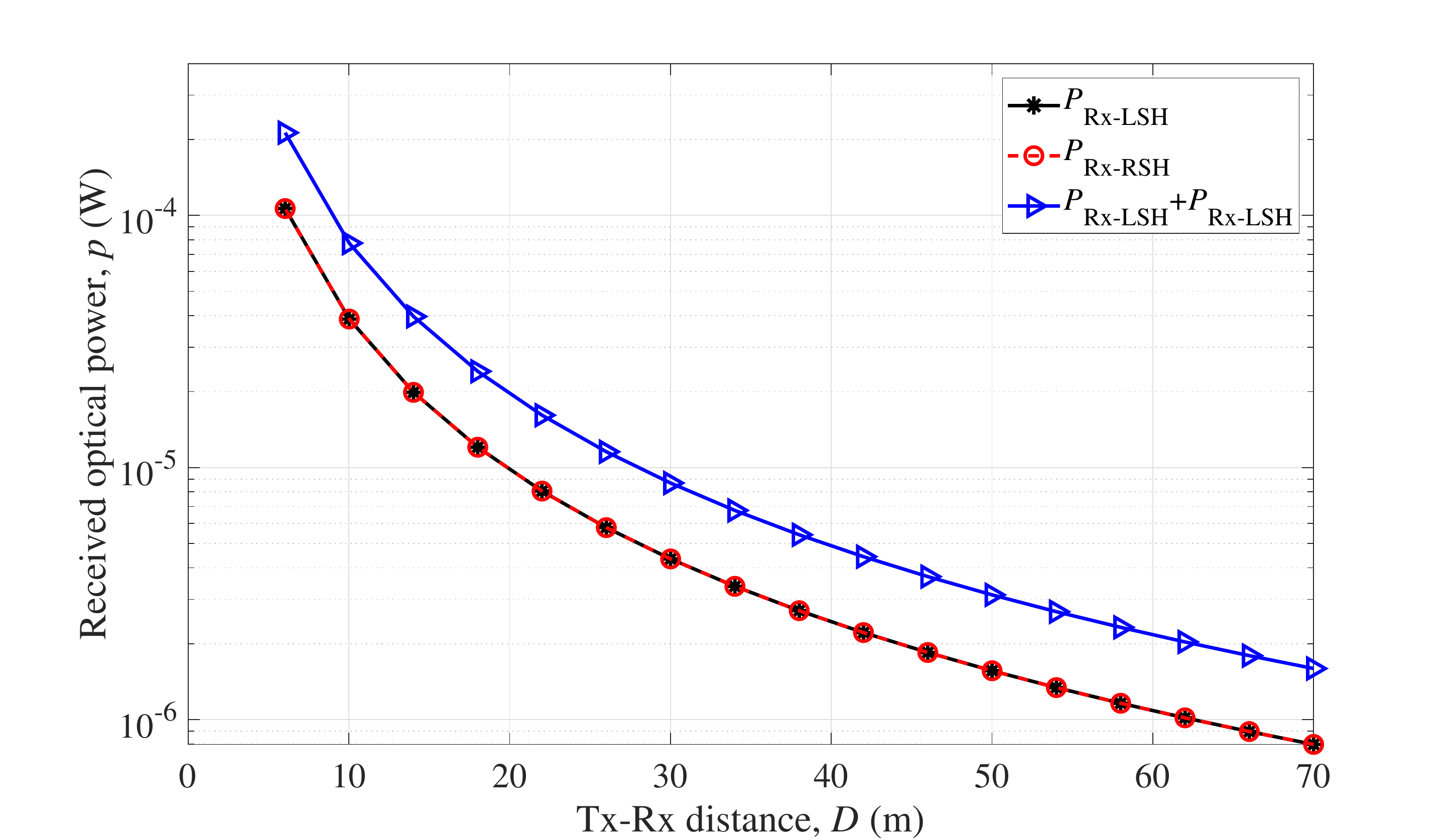} 
{Received power of LoS components vs. Tx-Rx distance ($\upsilon _{\mathrm{Tx}} = 21.6~\mathrm{km/h}$,~$\upsilon _{\mathrm{Rx}}=14.4~\mathrm{km/h} $, $\gamma_{\mathrm{T}}$ = $\gamma_{\mathrm{R}}$ = $0^{\circ}$, $2\delta= 1.2$ m, $\phi _{\mathrm{T}}=0^{\circ}$, $\mathrm{SD}=6$~m, $m$=1).\label{Fig_7}} 

\Figure[t!][scale=0.29]{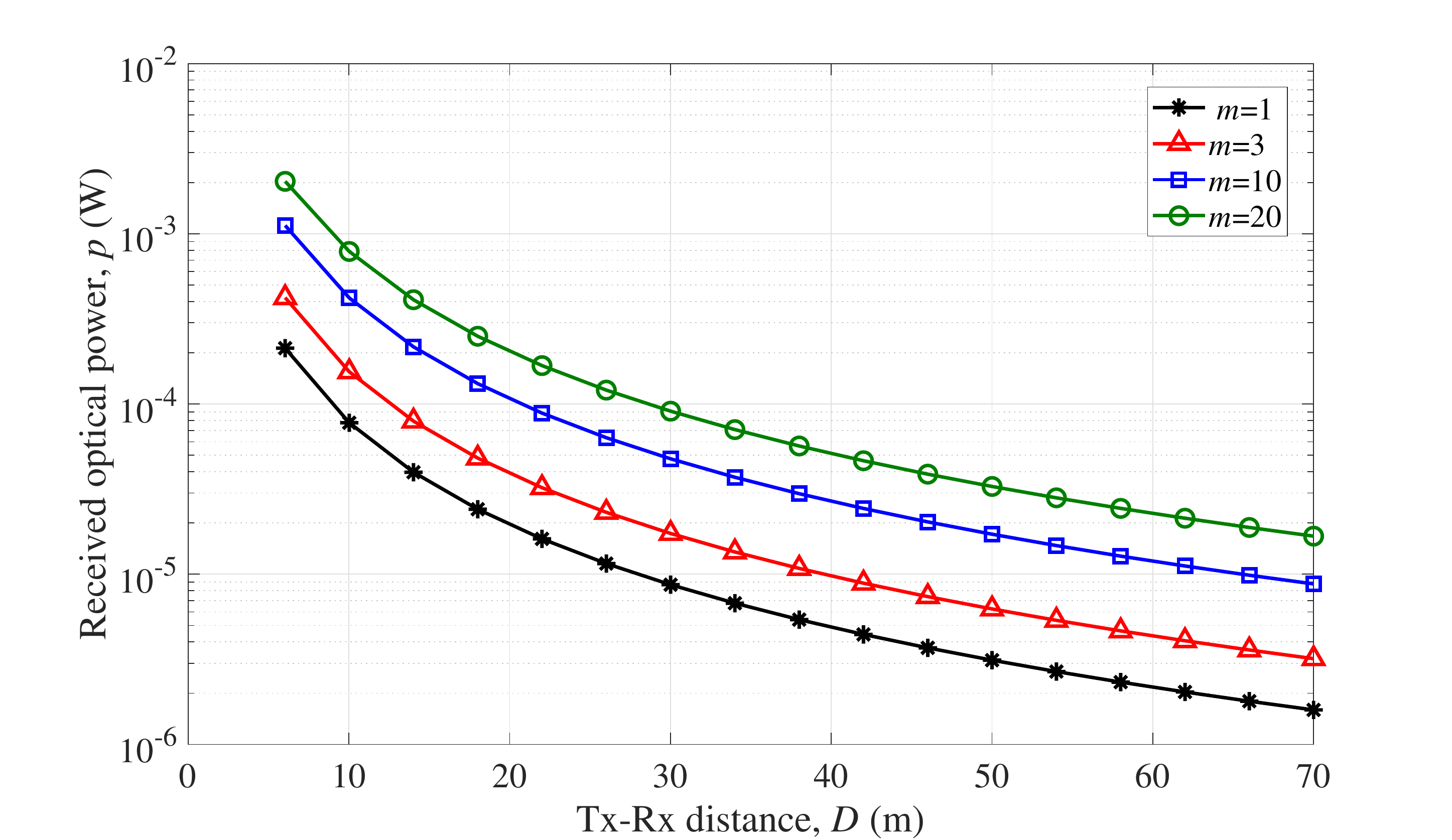} 
{Received power of LoS components vs. Tx-Rx distance ($\upsilon _{\mathrm{Tx}} = 21.6~\mathrm{km/h}$,~$\upsilon _{\mathrm{Rx}}=14.4~\mathrm{km/h} $, $\gamma_{\mathrm{T}}$ = $\gamma_{\mathrm{R}}$ = $0^{\circ}$, $2\delta= 1.2$ m, $\phi _{\mathrm{T}}=0^{\circ}$, $\mathrm{SD}=6$~m, $m$=1, 3, 10, 20).\label{Fig_8}}

\subsubsection{The SB components}
In this work, the VMF distribution have been adopted to take into account the joint impact of both azimuth and elevation angles on the channel characteristics. Since no measurements available, the main parameters including the mean values of the azimuth angles $\alpha_{0} ^{(i)}$ and elevation angles $\beta_{0}^{(i)}$ ($i = 1, 2, 3$), as well as the concentration factor $k$ will be assumed for simulation purposes. Accordingly, the related propagation distances can be determined by using the derived equations that presented in Sections~\ref{Section III} and \ref{Section_IV}. Here, we tried to make the assumptions as much as close to the reality. Since the effective scatterers are distributed according to the VMF distribution, we will investigate and analyze the effect of VMF parameters on the received power, namely, $N^{(i)}$, $k^{(i)}$, $\alpha^{(n_{i})}$, and~$\beta^{(n_{i})}$. For simulation purposes, appropriate values for the numbers of discrete scatterers $N^{(i)}$ must be chosen carefully~\cite{Pätzold12}. Based on our own simulation experiences, the value for $N^{(i)}$ is set to be 100. Furthermore, in order to obtain the set of~$\left \{\alpha^{(n_{i})},\beta^{(n_{i})} \right \}^{N_{i}}_{n_{i}=1}$ we use the method of equal volume~(MEV), which is proposed in~\cite{Yuan14} to generate discrete values for the azimuth and elevation angles around the mean direction. In the following subsections, the effect of each parameter on the received power is studied separately for each model.\par

\paragraph {Received Optical Power From Tx-Sphere and Rx-Sphere Models}
The received power is related to the concentration factor $k$, which is managing the distribution of the scatterers according to VMF distribution. In this paper, we examine the impact of $k$ and the mean direction of the scatterers on the power amount that can be received from LSH and RSH. Here, the mean direction of the scatterers indicates the position of a car in the adjacent lane. In reality, vehicles are not aligned precisely with other surrounding vehicles that located in the adjacent lanes. Hence, cars which are on the right side make different angles compared with the cars on the left side. In order to consider the cars on the left side and right sides, two sets of mean angles have been defined, $\left \{\alpha_{0}^{(n_{i},\mathrm{L})},\beta_{0}^{(n_{i},\mathrm{L})} \right \}$ and $\left \{\alpha_{0}^{(n_{i},\mathrm{R})},\beta_{0}^{(n_{i},\mathrm{R})}\right \}$, respectively. As VVLC technology is still growing and at an early stage of research, there are currently no available measurements data. Therefore, we tried to take into account the most reliable parameters which are as close to reality as possible and hence we set the value of the elevation angle to $2^{\circ}$. This is due to the fact that in urban environments, the majority of cars will be sedan cars and hence the reflection from the surrounding cars will be at almost the same plane. On the other hand, the azimuth angles have been set to values as $\alpha_{0}^{(n_{i},\mathrm{L})}=\alpha_{0}^{(n_{i},\mathrm{R})}=10^{\circ}, 30^{\circ}, 45^{\circ}$. The azimuth angles were chosen to ensure that the most probable positions for the adjacent cars have been taken into account. Furthermore, regarding~$k_{\mathrm{c}}^{(i)}$, we followed the procedure that used in conventional RF V2V scenarios in~\cite{Yuan14}. However, for VVLC, the value of $k_{\mathrm{c}}^{(i)}$ has been set to 3, 14, and 30. The other parameters which are related to the proposed model are listed in Table~\ref{Model Parameters_3D}.
Since the LSH and RSH present the same behavior, here only the powers which are generated at LSH and reflected off the surrounding vehicles on the left side will be analyzed. Fig.~\ref{Fig_9} and Fig.~\ref{Fig_10} illustrate the received power from LSH within Tx-Sphere model and Rx-Sphere model, respectively. 
\Figure[t!][scale=0.29]{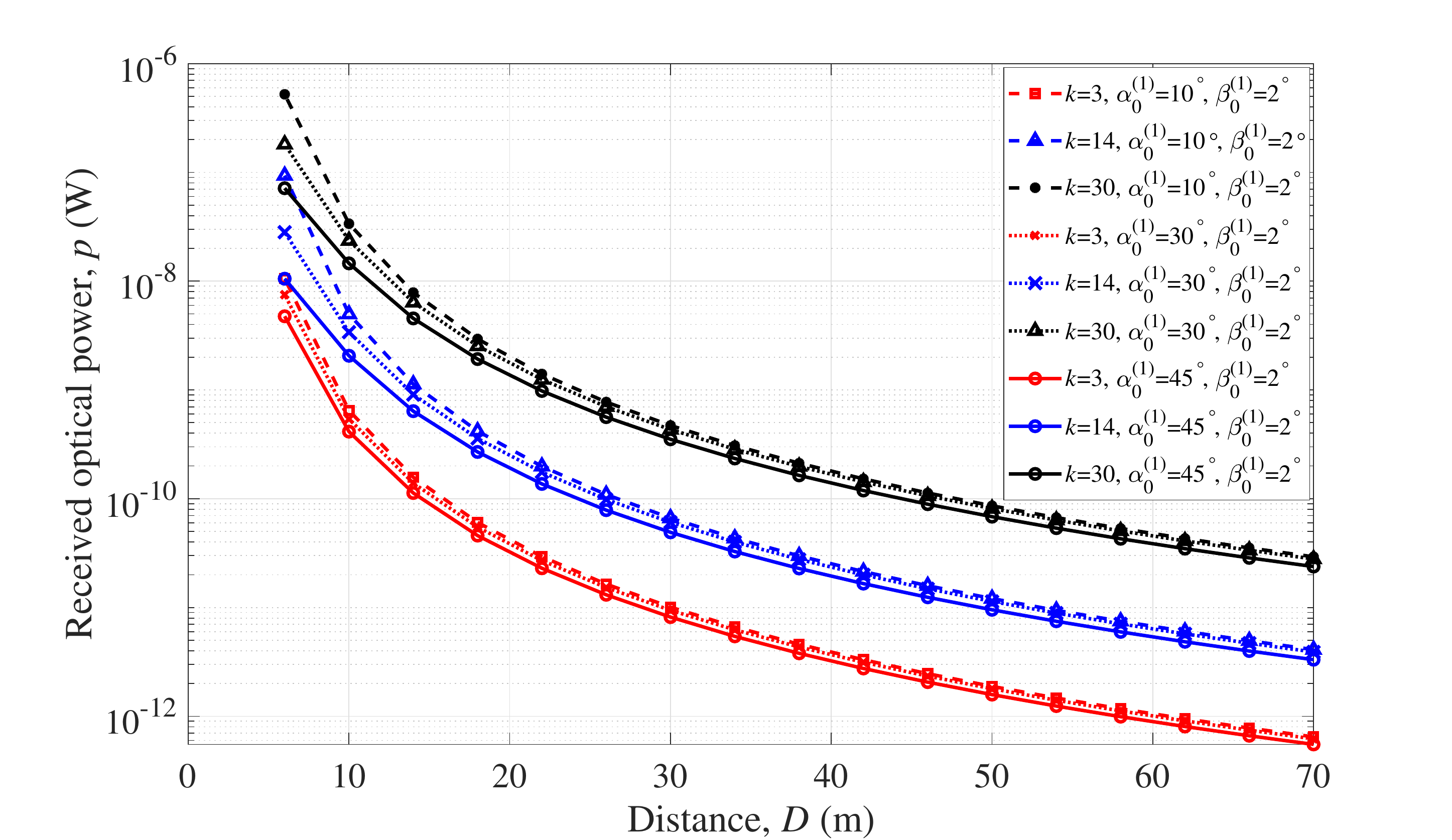} 
{Received power from LSH within Tx-Sphere model ($\gamma_{T}= \gamma_{R}=0^{\circ}$, $\delta=0.6$~m, $\phi _{\mathrm{T}}=0$, $\alpha_{0}^{(1)}=10^{\circ}, 30^{\circ}, 45^{\circ}$, $\beta_{0}^{(1)}=2^{\circ}$, $k=3, 10, 30$).\label{Fig_9}}
\Figure[t!][scale=0.29]{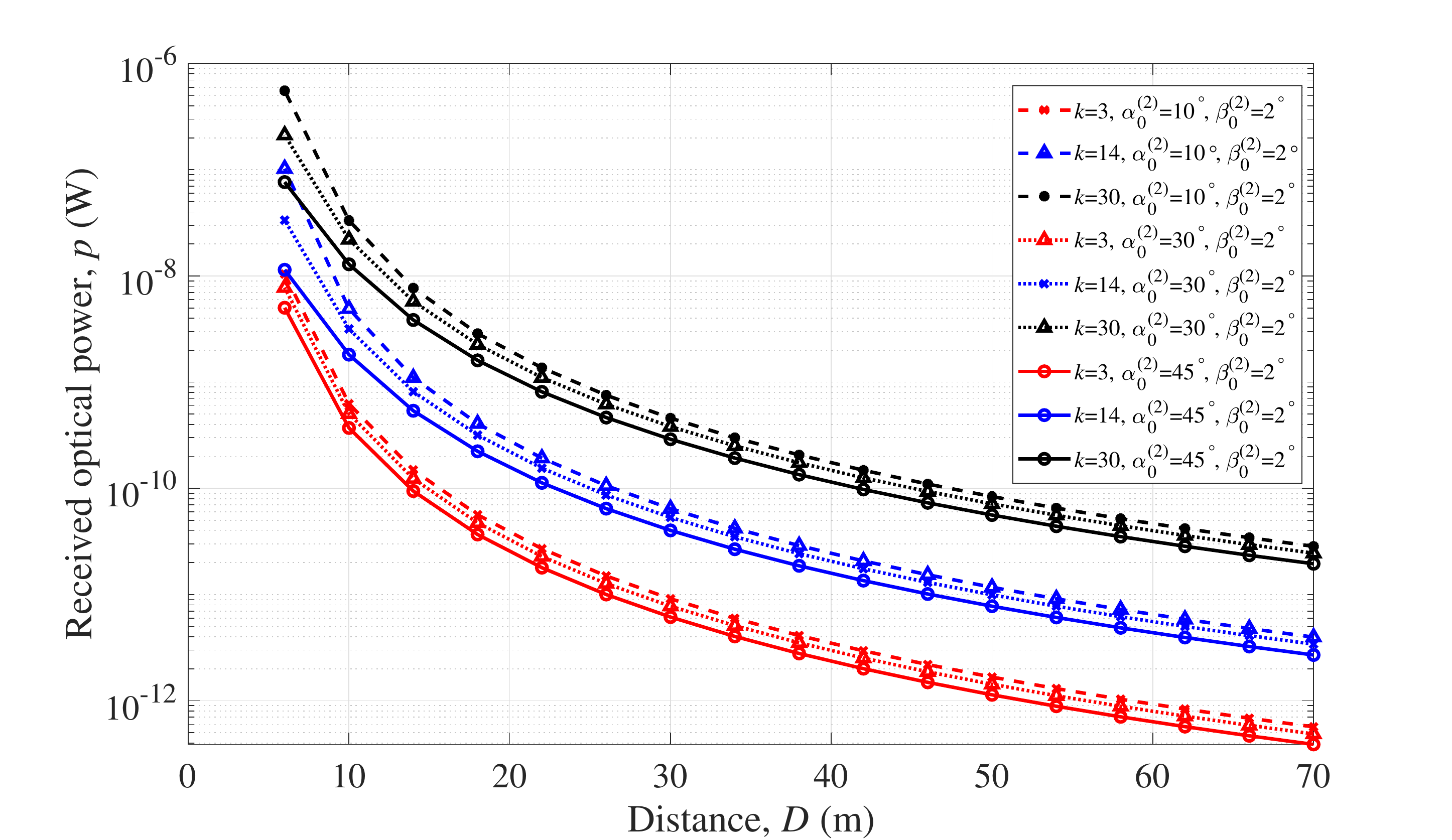} 
{Received power from LSH within Rx-Sphere model ($\gamma_{T}= \gamma_{R}= 0^{\circ}$, $\delta= 0.6$~m, $\phi _{\mathrm{T}}= 0$, $\alpha_{0}^{(2)}= 10^{\circ}, 30^{\circ}, 45^{\circ}$, $\beta_{0}^{(2)}=2^{\circ}$, $k=3, 10, 30$).\label{Fig_10}}
It can be realized that higher optical power will be received as $k$ goes higher. This is due that the higher $k$ means the local scatterers are being highly aligned around the mean angles. On the other hand, when the mean angles increase, the received power decrease. For instance, in Fig.~\ref{Fig_9}, for $k=30$ the received power from the LSH that reflected off an obstacle located at the left side at a distance of 10~m is $3.37\times 10^{-8}$~W when the mean angle $\alpha_{0}^{(1)}= 10^{\circ}$. While the received power is $1.45\times 10^{-8}$~W when $\alpha_{0}^{(1)}= 45^{\circ}$, for the same $k$ value. This is due to Lambert's cosine law as the light intensity is correlated to the angle with surface normal of the LED headlight and PD. Furthermore, as much as $k$ decreases, there will be no dominant mean direction and hence more deviation about the surface normal.\par
It is worth mentioning that in terms of Rx-sphere model, the FoV constraint of the PD is considered. Consequently, the assumed mean angles, i.e.,~$\alpha_{0}^{(2)}$, and $\beta_{0}^{(2)}$ must be within PD's FoV $\mathrm{\Psi _{FoV}}$.

\paragraph {Received Optical Power From Elliptic-Cylinder Model}
In terms of the elliptic-cylinder model, it is intuitive that less power will be received compared with the two-sphere model. This is due to two main reasons, firstly, the lower reflectivity of the roadside environments ${\rho }_{\mathrm{Roadside}}$. Secondly, the longer optical path lengths within the elliptic-cylinder model. In this case, the same behavior which is noticed in the Tx- and Rx-sphere appears here so that higher $k$ produces much power at the PD. Fig.~\ref{Fig_11} illustrates the received optical power, which is transmitted from LSH then reflected off the roadside environments to be detected by the PD.
For SB components, it can be noted from the above figures that at distances shorter than 10~m, the model cannot simulate satisfactorily the behavior of the wireless optical channel since some results are overlapped. This is due the radius of the Tx and Rx spheres, since no applicable reflection can be occurred at the range of 8~m.
\Figure[t!][scale=0.29]{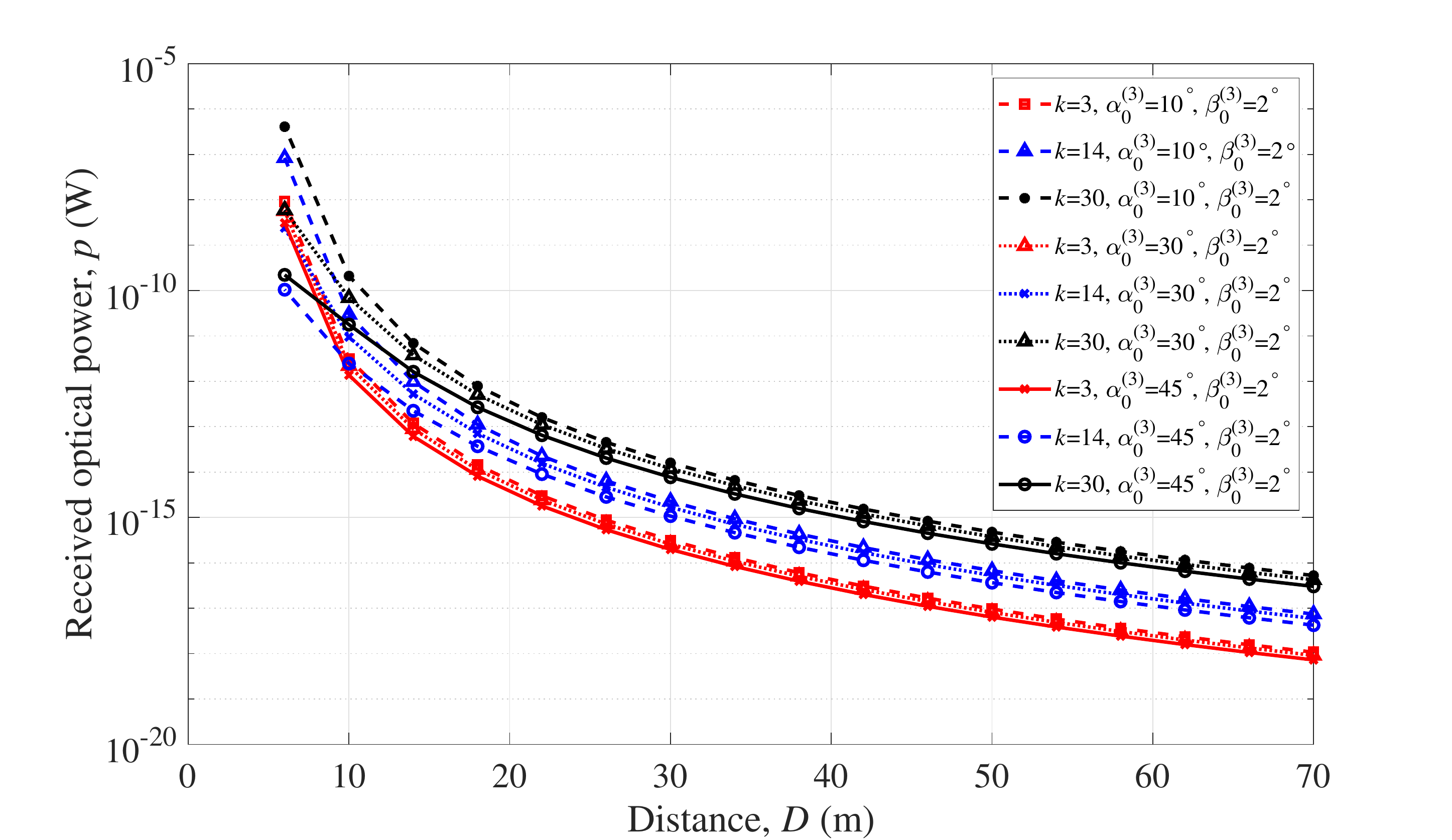} 
{Received power from LSH within elliptic-cylinder model ($\gamma_{T} = \gamma_{R}= 0^{\circ}$, $\delta= 0.6$~m, $\phi _{\mathrm{T}}= 0$, $\alpha_{0}^{(3)}= 10^{\circ}, 30^{\circ}, 45^{\circ}$, $\beta_{0}^{(3)}=2^{\circ}$, $k=3, 10, 30$).\label{Fig_11}}

On the other hand, in order to show the amount of power added by the SB components, Fig.~\ref{Fig_12} illustrates the LoS power in addition to SB power of the LSH. It can be seen that the LoS component plays a decisive role in the received optical power and the SB components add insignificant amounts of added powers, especially at longer Tx-Rx distances. In addition, the power difference between the LoS component and SB component increases as the Tx-Rx distance increases.
\Figure[t!][scale=0.29]{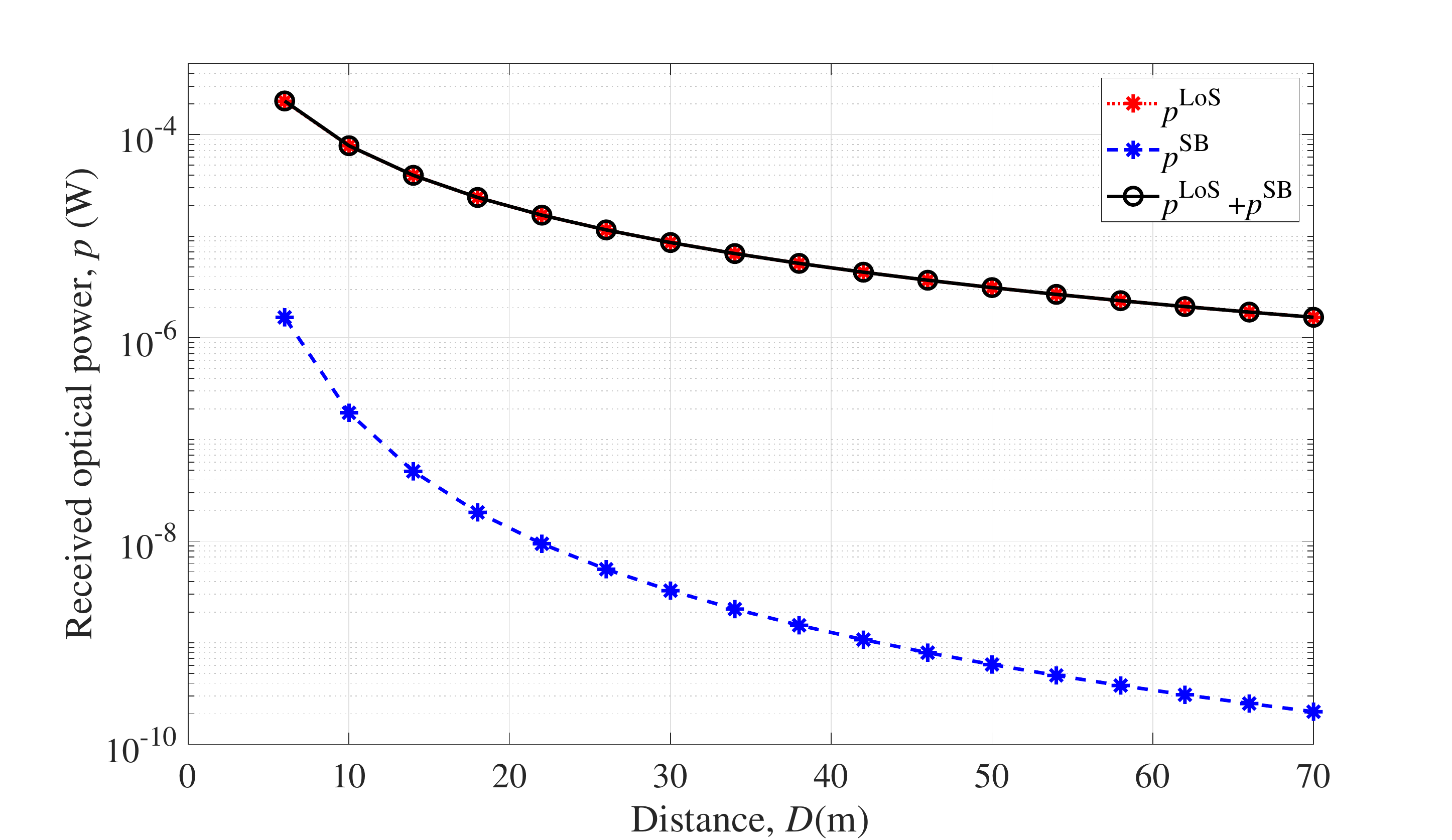} 
{The received power of LoS and SB components of LSH ($k=30, \alpha_{0}^{(i)}= 30^{\circ},~\beta_{0}^{(i)},=2^{\circ},  i=1, 2, 3$).\label{Fig_12}}

\Figure[t!][scale=0.29]{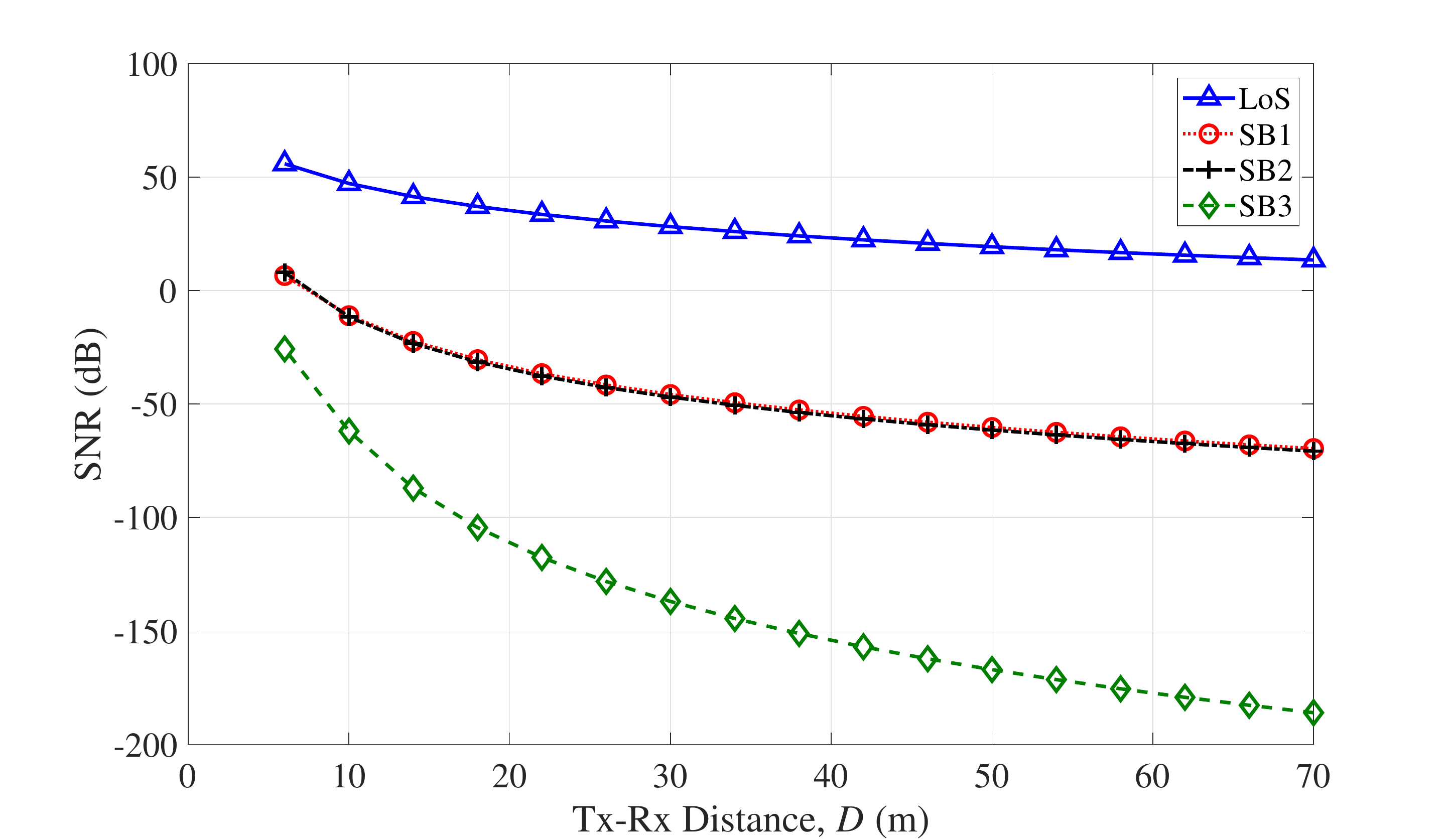} 
{SNR vs. Tx-Rx distance ($ k=30, \alpha_{0}^{(i)}= 10^{\circ},~\beta_{0}^{(i)},=2^{\circ}, i=1, 2, 3$).\label{Fig_13}}

\subsection{SNR}
Based on the noise analysis in Section~\ref{Section_IV}$-\mathrm{C}$, the performance of each component can be analyzed through the relationship between the SNR and Tx-Rx distance as illustrated in Fig.~\ref{Fig_13}. Here, only the assumption of $\alpha_{0}^{(i)}= 10^{\circ}$, $\beta_{0}^{(i)}=2^{\circ}$, and $k=30$ has been considered for all components. This is due to the fact that these components carry higher power compared with the others. It can be noticed from Fig.~\ref{Fig_13}, that SNR values decrease as the Tx-Rx distances increase and the difference in received power according to each component is maintained.

\subsection{Comparison Between 3D VVLC RS-GBSM and 2D VVLC RS-GBSM}
Regarding the proposed 3D RS-GBSM, it is worth to emphasize that when $\beta^{(n_{i})}=0$,~$(i=1,2,3)$, the proposed 3D model will be reduced to a 2D RS-GBSM (two-ring and elliptic model) in~\cite{2D SISO}. In order to evaluate the impact of elevation angle on the received power, Fig.~\ref{Fig_14} and Fig.\ref{Fig_15} illustrate comparisons between the 3D and 2D models in terms of the received power from LSH for the LoS and SB components, respectively. Note that we considered the same number of effective scatterers, i.e., $N=100$. From Fig.\ref{Fig_14} and Fig.\ref{Fig_15}, it is clear that compared with the 3D model, the 2D model overestimates the received optical power. The reason is that the 2D model assumes that $\beta^{(n_{i})}$ has no contribution. Moreover, compared to the 2D model in~\cite{2D SISO}, 3D model introduces an extra optical path length caused by considering the headlight separation~$2\delta$.
\Figure[t!][scale=0.29]{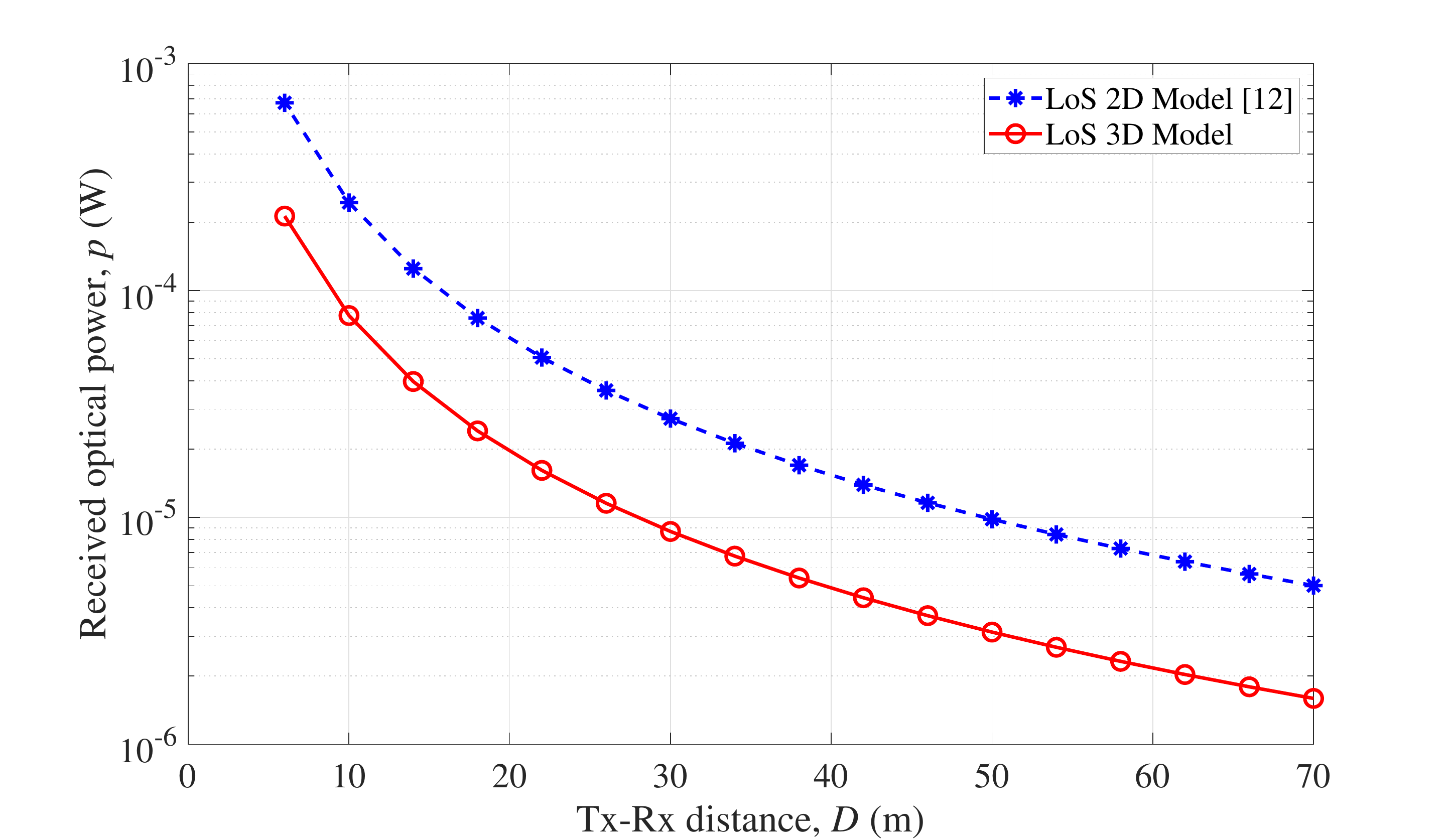} 
{LoS received power comparison between the 3D and 2D models ($\upsilon _{\mathrm{Tx}} = 21.6~\mathrm{km/h}$,~$\upsilon _{\mathrm{Rx}}=14.4~\mathrm{km/h} $, $\gamma_{\mathrm{T}}$ = $\gamma_{\mathrm{R}}$ = $0^{\circ}$, $2\delta= 1.2$ m, $\phi _{\mathrm{T}}=0^{\circ}$, $m$=1).\label{Fig_14}}

\Figure[t!][scale=0.29]{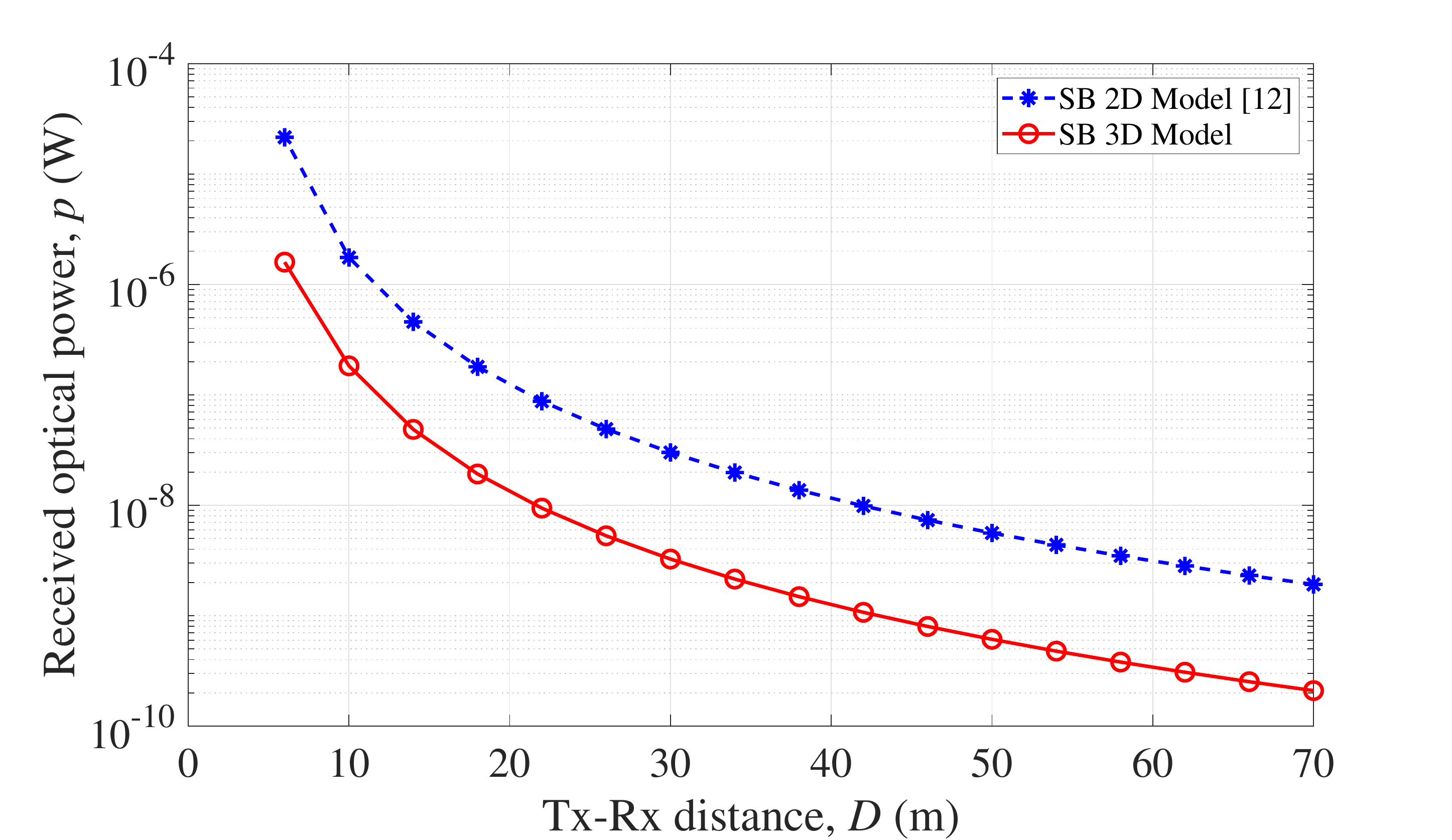} 
{SB received power comparison between the 3D and 2D models ($\upsilon _{\mathrm{Tx}} = 21.6~\mathrm{km/h}$,~$\upsilon _{\mathrm{Rx}}=14.4~\mathrm{km/h} $, $\gamma_{\mathrm{T}}$ = $\gamma_{\mathrm{R}}$ = $0^{\circ}$, $2\delta= 1.2$ m, $\phi _{\mathrm{T}}=0^{\circ}$, $m$=1).\label{Fig_15}}
\section{Conclusions}\label{Section_VI}
In this paper, a new 3D non-stationary RS-GBSM for VVLC MISO channels has been proposed. The proposed model jointly considers the azimuth and elevation angles by using the VMF distribution. VVLC channel characteristics have been examined through a large set of channel impulse responses generated by the proposed 3D RS-GBSM. The received optical powers for the LoS and SB components have been computed along different distance ranges between 0 and 70 m. Simulation results have shown that for the proposed model, the azimuth angle has a significant impact on the received power. This is due to the fact that light intensity is correlated with the cosine of the observation angle with respect to the surface normal of the LED headlight and PD. Moreover, the background noise sources have been modeled and the VVLC system's SNR has been evaluated accordingly. Finally, it has been demonstrated that compared with the 3D model, the 2D model overestimates the received optical~power.

\begin{IEEEbiography}[{\includegraphics[width=1in,height=1.25in,clip,keepaspectratio]{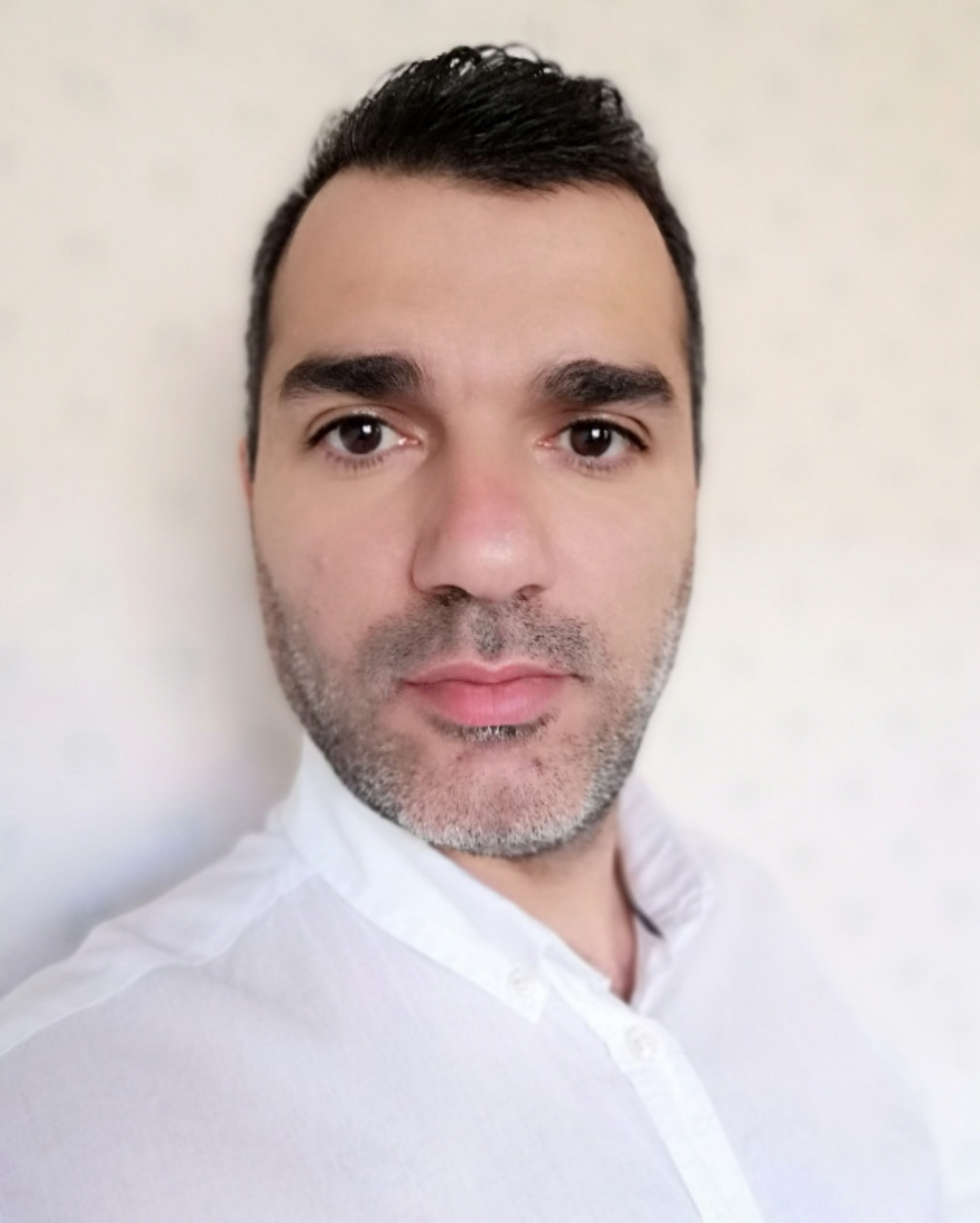}}]{AHMED AL-KINANI} received the B.Sc. degree (Hons.) in Laser technology  and the M.Sc. degree in optical communications from the University of Technology, Baghdad, Iraq, in 2001 and 2004, respectively, and the Ph.D. degrees in optical wireless communications from the University of Edinburgh and Heriot-Watt University, Edinburgh, U.K., in 2018. He was a recipient of the Best Paper Awards from IEEE IWCMC 2016. He is currently with Cellular Asset Management Services as a radio engineer. His research interests include optical wireless communications, optical channel characterization and modeling for visible light communications, and heterogeneous 5G and beyond networks.
\end{IEEEbiography}

\begin{IEEEbiography}[{\includegraphics[width=1in,height=1.25in,clip,keepaspectratio]{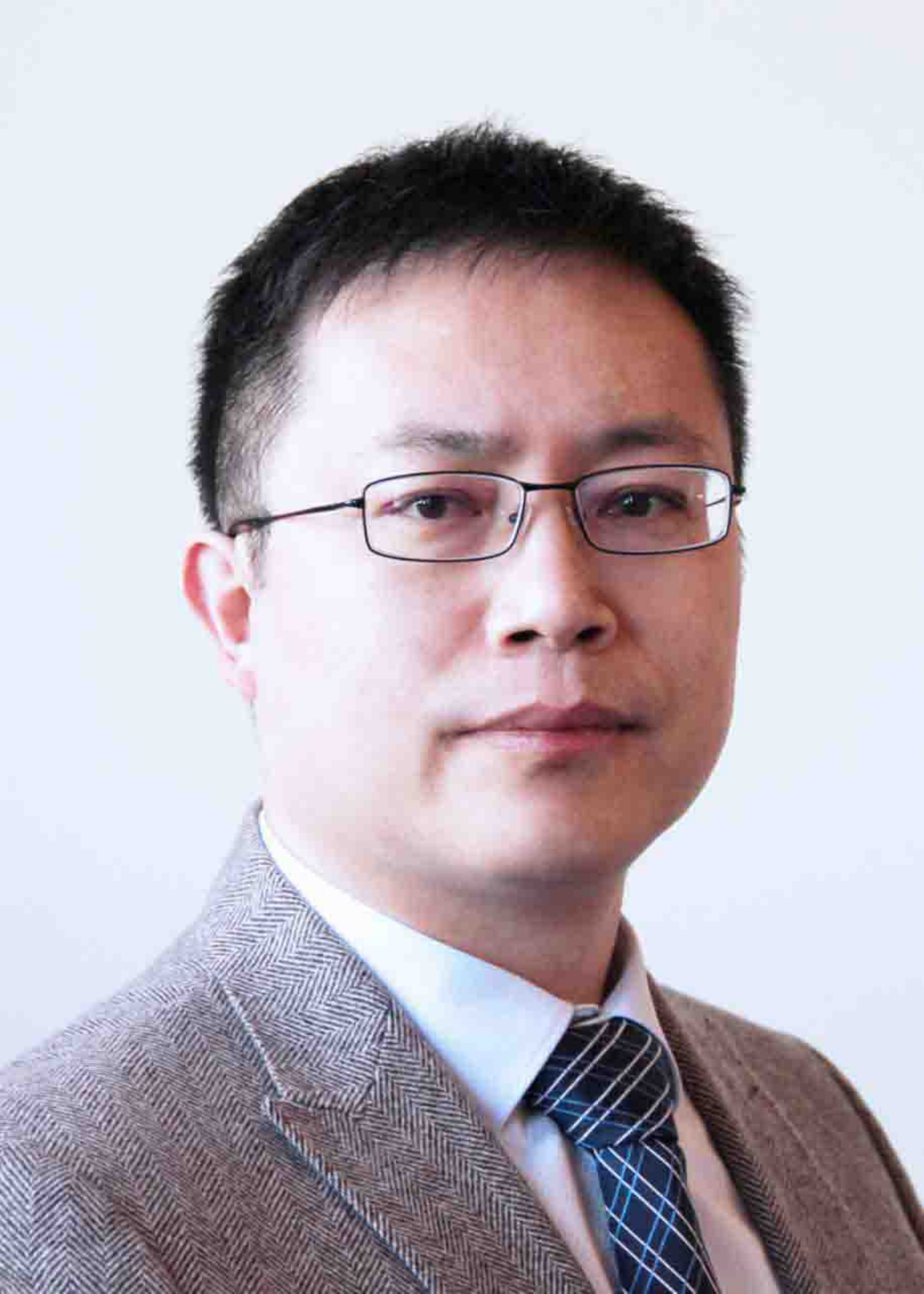}}]{CHENG-XIANG WANG} (S'01-M'05-SM'08-F'17) received the B.Sc. and M.Eng. degrees in communication and information systems from Shandong University, China, in 1997 and 2000,
respectively, and the Ph.D. degree in wireless communications from Aalborg University, Denmark, in 2004. 

He was a Research Assistant with the Hamburg University of Technology, Hamburg, Germany, from 2000 to 2001, a Visiting Researcher with Siemens AG Mobile Phones, Munich, Germany, in 2004, and a Research Fellow with the University of Agder, Grimstad, Norway, from 2001 to 2005. He has been with Heriot-Watt University, Edinburgh, U.K., since 2005, where he was promoted to a Professor, in 2011. In 2018, he joined Southeast University, China, as a Professor. He is also a part-time professor with the Purple Mountain Laboratories, Nanjing, China. He has authored four books, two book chapters, and over 390 papers in refereed journals and conference proceedings, including 23 Highly Cited Papers. He has also delivered 20 Invited Keynote Speeches/Talks and seven Tutorials in international
conferences. His current research interests include wireless channel measurements and modeling, B5G wireless communication networks, and
applying artificial intelligence to wireless communication networks. 

Dr. Wang is a Fellow of the IET, an IEEE Communications Society
Distinguished Lecturer in 2019 and 2020, and a Highly-Cited Researcher recognized by Clarivate Analytics in 2017-2019. He is currently an Executive
Editorial Committee Member of the IEEE TRANSACTIONS ON WIRELESS COMMUNICATIONS. He has served as an Editor for nine international journals, including the IEEE TRANSACTIONS ONWIRELESS COMMUNICATIONS, from 2007 to 2009, the IEEE TRANSACTIONS ON VEHICULAR TECHNOLOGY, from 2011 to 2017, and the IEEE TRANSACTIONS ON COMMUNICATIONS, from 2015 to 2017. He was a Guest Editor of the IEEE JOURNAL ON SELECTED AREAS IN COMMUNICATIONS, Special Issue on Vehicular Communications and Networks (Lead Guest Editor), Special Issue on Spectrum and Energy Efficient Design of Wireless Communication Networks, and Special Issue on Airborne Communication Networks. He was also a Guest Editor for the IEEE TRANSACTIONS ON BIG DATA, Special Issue on Wireless Big Data, and is a Guest Editor for the IEEE TRANSACTIONS ON COGNITIVE COMMUNICATIONS AND NETWORKING, Special Issue on Intelligent Resource Management for 5G and Beyond. He has served as a TPC Member, a TPC Chair, and a General Chair for more than 80
international conferences. He received eleven Best Paper Awards from IEEE GLOBECOM 2010, IEEE ICCT 2011, ITST 2012, IEEE VTC 2013-Spring, IWCMC 2015, IWCMC 2016, IEEE/CIC ICCC 2016, WPMC 2016, WOCC 2019, and IWCMC 2020.
\end{IEEEbiography}

\begin{IEEEbiography}[{\includegraphics[width=1in,height=1.25in,clip,keepaspectratio]{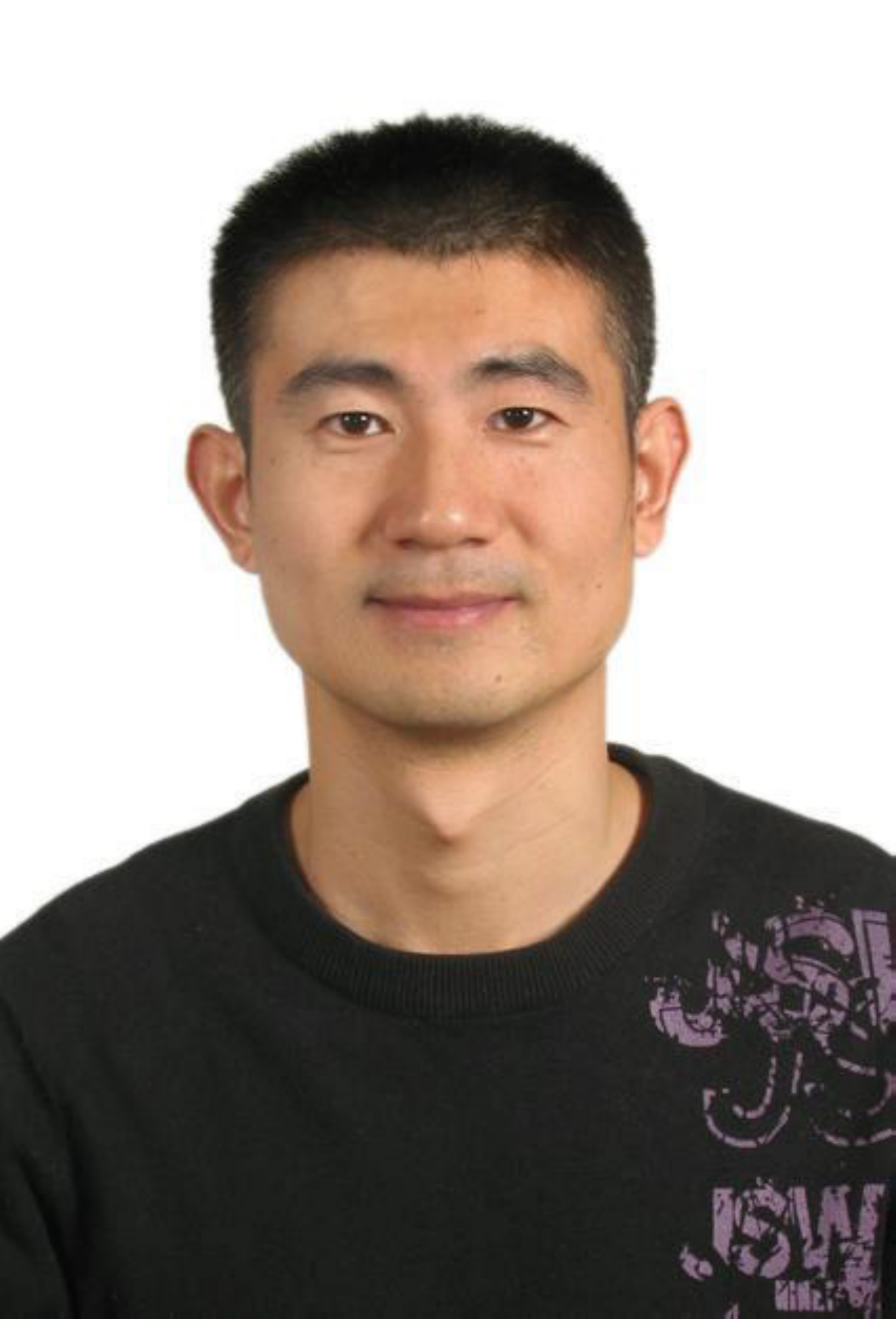}}]{QIUMING ZHU} received the B.Sc. degree in electronic engineering from Nanjing University of Aeronautics and Astronautics (NUAA) in Nanjing, China in 2002 and his M.Sc. and Ph.D. degrees in communication and information systems in 2005 and 2012, respectively. Since 2012, he has been an associate professor in wireless communications. From 2016 to 2017, he was also a visiting academic at Heriot-Watt University. His research interests include channel modeling for 5G communication systems and wireless channel emulators.
\end{IEEEbiography}

\begin{IEEEbiography}[{\includegraphics[width=1in,height=1.25in,clip,keepaspectratio]{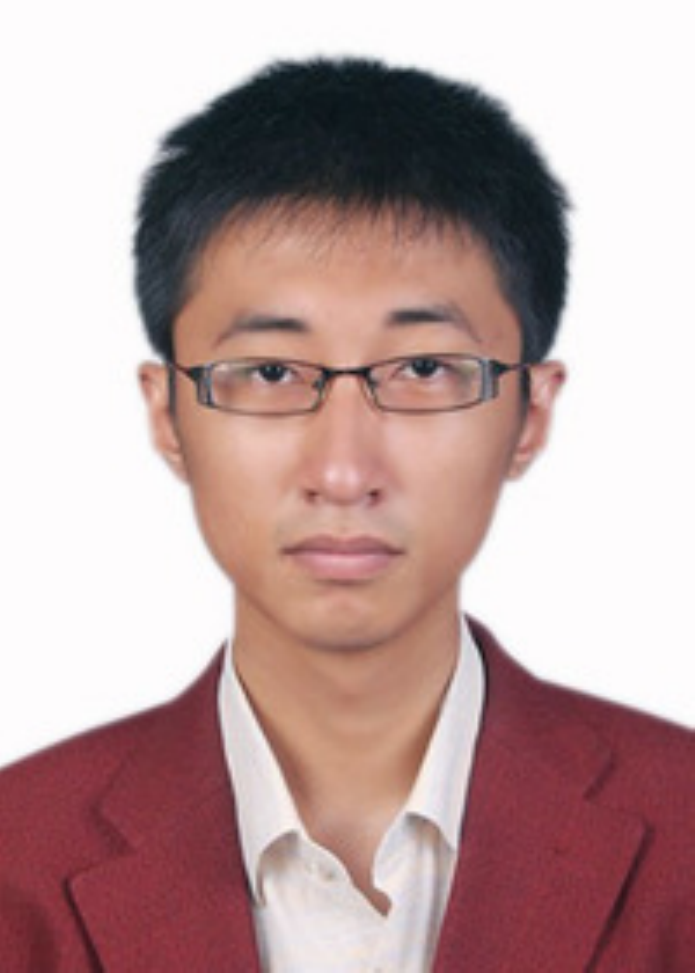}}]{YU FU} received his BSc degree in Computer Science from Huaqiao University, Fujian, China, in 2009, the MSc degree in Information Technology (Mobile Communications) from Heriot-Watt University, Edinburgh, U.K., in 2010, and the PhD degree in Wireless Communications from Heriot-Watt University, Edinburgh, UK, in 2015. He has been a Postdoc Research Associate of Heriot-Watt University since 2015. His main research interests include advanced MIMO communication technologies, wireless channel modeling and simulation, RF tests, and software defined networks.
\end{IEEEbiography}

\begin{IEEEbiography}[{\includegraphics[width=1in,height=1.25in,clip,keepaspectratio]{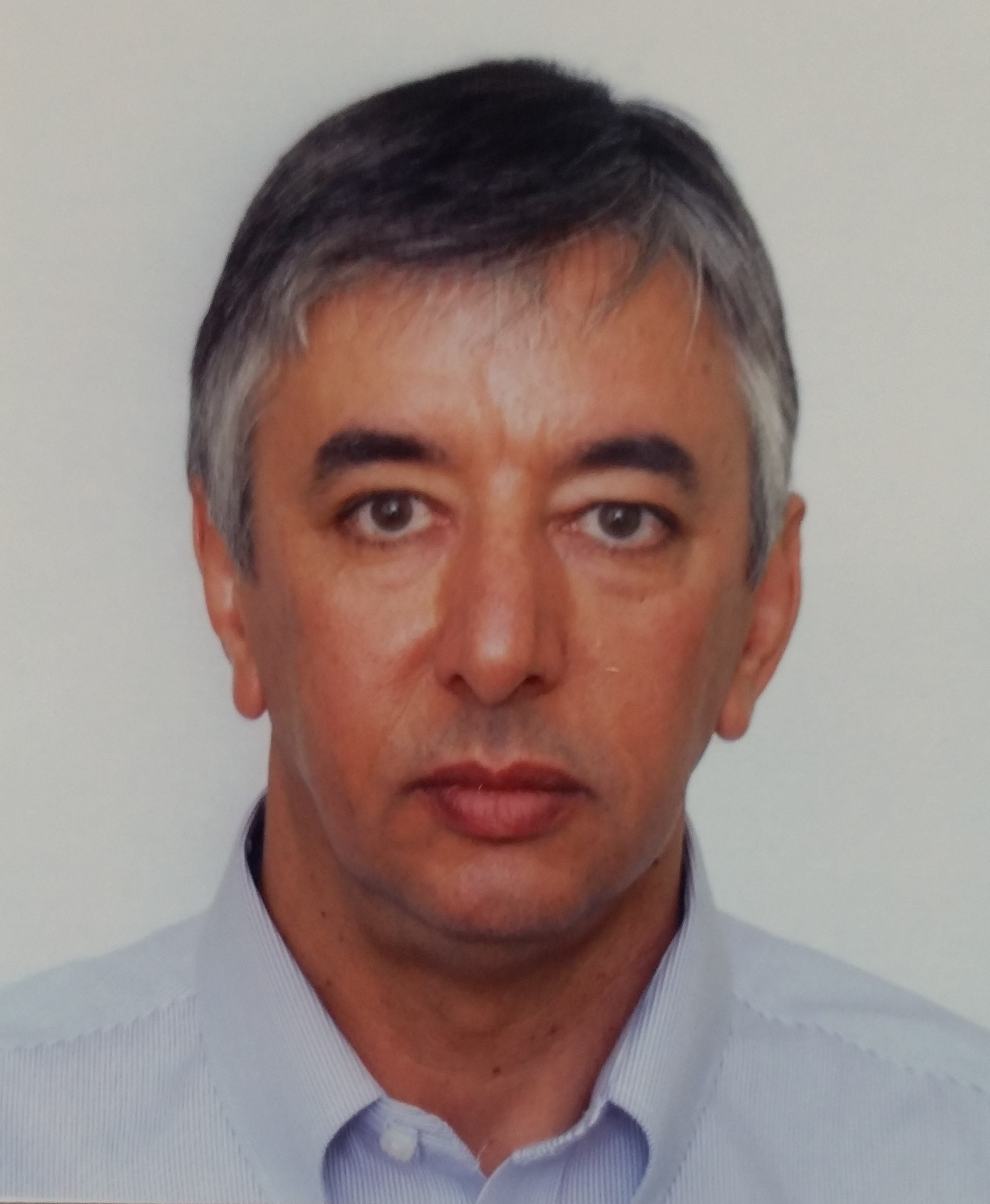}}]{EL-HADI M. AGGOUNE} (M'83-SM'93-SLM'20) received the M.S. and Ph.D. degrees in electrical engineering from the University of Washington (UW), Seattle, WA, USA. He taught graduate and undergraduate courses in electrical engineering at many universities in the USA and abroad. He served at many academic ranks, including an Endowed Chair Professor. He is listed as an Inventor in two patents assigned to the Boeing Company, USA, and the Sensor Networks and Cellular Systems Research Center, University of Tabuk, Saudi Arabia. He is also a Professional Engineer registered in Washington. He is currently serving as a Professor and the Director of the SNCS Research Center, University of Tabuk. His research is referred to in many patents, including patents assigned to ABB, Switzerland, and EPRI, USA. He has authored many papers in IEEE and other journals and conferences. His research interests include wireless sensor networks, energy systems, and scientific visualization. He is serving on many technical committees for conferences worldwide and a Reviewer for many journals. One of his Laboratories received the Boeing Supplier Excellence Award. He was the recipient of the IEEE Professor of the Year Award, UW Branch.
\end{IEEEbiography}

\begin{IEEEbiography}[{\includegraphics[width=1in,height=1.25in,clip,keepaspectratio]{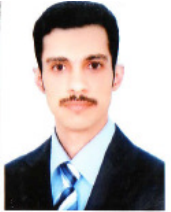}}]{AHMED TALIB} received the B.Sc. degree from the College of electrical and electronic techniques, Baghdad, Iraq, in 2004, and the M.Sc. degree from Brunel University, London, UK, in 2013. He is currently with Iraqi Telecommunication and Informatics Company (ITPC), Iraq, as a switching department manager. His research interests include 5G and beyond technologies.
\end{IEEEbiography}

\begin{IEEEbiography}[{\includegraphics[width=1in,height=1.25in,clip,keepaspectratio]{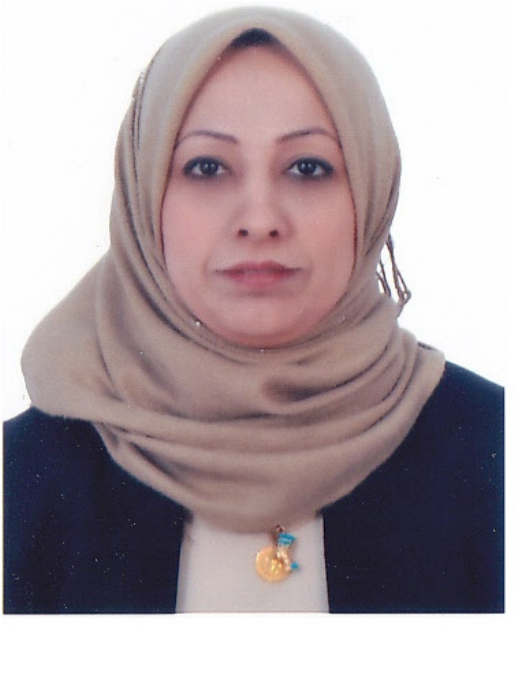}}]{NIDAA AL-HASAANI} received the B. Sc degree in electrical engineering from Baghdad University, Iraq, in 1995, and the M. Sc degree in electronics and communication engineering from Yildiz Technicall University, Istanbul, Turkey in 2018. From 2005 to 2008, she joined Zain Telecom, Iraq, as a RAN manager. She is currently with Iraqi Telecommunication and Informatics Company (ITPC), Iraq, as chief senior engineer, Iraq. Her research includes GFDM and OFDM modulations, spectrum sensing, cognitive radio, energy detection, and channel estimation.
\end{IEEEbiography}

 \EOD


\begin{thebibliography}{00}
\bibitem{WHO 2016}
 World Health Organization. Accessed: Mar. 30, 2020. [Online]. Available: http://www.who.int/mediacentre/factsheets/fs358/en/.

\bibitem{Zeadally2012}
S. Zeadally, R. Hunt, Y.-S. Chen, A. Irwin, and A. Hassan, ``Vehicular ad hoc networks (VANETS): status, results, and challenges,'' {\it Telecom. Syst.}, vol. 50, no. 4, pp. 217--241, Aug. 2012.

\bibitem{Fourat14} 
C.-X. Wang, F. Haider, X. Gao, X.-H. You, Y. Yang, D. Yuan, H. Aggoune, H. Haas, S. Fletcher, and E. Hepsaydir, ``Cellular architecture and key technologies for 5G wireless communication networks,'' {\it IEEE Commun. Mag.}, vol. 52, no. 2, pp. 122--130, Feb. 2014.

\bibitem{Thesis}
A. Al-Kinani, ``Channel modelling for visible light communication systems," Ph. D. thesis, Heriot-Watt University, Edinburgh, UK, 2018. [Online]. Available: https://www.ros.hw.ac.uk/handle/10399/3486.

\bibitem{A.Al-Kinani} 
A. Al-Kinani, C.-X. Wang, H. Haas, and Y. Yang, ``A geometry-based multiple bounce model for visible light communication channels,'' in {\it Proc. IEEE IWCMC'16}, Paphos, Cyprus, Sept. 2016, pp. 31--37.


\bibitem{Wada 05}
M. Wada, T. Yendo, T. Fujii, and M. Tanimoto, ``Road-to-vehicle communication using LED traffic light,'' in {\it IEEE Intell. Veh. Symp.}, Nevada, USA, June 2005, pp. 601--606.

\bibitem{Pang 02}
G. Pang, T. Kwan, H. Liu, and C.-H Chan, ``LED wireless,''  {\it IEEE Industry Applications Mag.}, vol. 8, no. 1, pp. 21--28, Jan. 2002.

\bibitem{Iwasaki2008}
 S. Iwasaki, C. Premachandra, T. Endo, T. Fujii, M. Tanimoto, and Y. Kimura, ``Visible light road-to-vehicle communication using high-speed camera,'' in {\it Proc. IEEE Intell. Veh. Symp.}, Eindhoven, The Netherlands, Jun. 2008, pp. 13--18.


\bibitem{Lee2012}
S. J. Lee, J. K. Kwon, S. Y. Jung, and Y. H. Kwon, ``Simulation modelling of visible light communication channel for automotive applications,'' in {\it Proc. IEEE ITSC'12}, Anchorage, USA, Sept. 2012, pp.~463--468.

\bibitem{S. Lee2012}
S. J. Lee, J. K. Kwon, S. Y. Jung, and Y. H. Kwon, ``Evaluation of visible light communication channel delay profiles for automotive applications,'' {\it EURASIP J. Wirel. Commun. Netw.}, pp.1--8, 2012.

\bibitem{AhmedS}
A. Al-Kinani, C.-X. Wang, L. Zhou, and W. Zhang, ``Optical wireless communication channel measurements and models,'' {\it IEEE Commun. Surveys Tuts.}, vol. 20, no.3, pp. 1939--1962, 3rd Quart., 2018.

\bibitem{Luo2015}
P. Luo, Z. Ghassemlooy, H.- L. Minh, E. Bentley, A, Burton, and X. Tang, ``Performance analysis of a car-to-car visible light communication system,'' {\it Appl. Opt.}, vol. 54, no. 7, pp. 1696--1706, Mar. 2015.

\bibitem{I-Table}
B. Schoettle and M. J. Flannagan, ``A Market-Weighted Description of Low-Beam and High-Beam Headlighting Patterns in the U.S.,'' University of Michigan, Michigan U.S.A. Rep. UMTRI-2004-23, 2004.

\bibitem{2D SISO}
A. Al-Kinani, J. Sun, C.-X. Wang, W. Zhang, X. Ge, and H. Haas, ``A 2D non-stationary GBSM for vehicular visible light communication channels,'' {\it IEEE Trans. Wireless Commun.}, vol. 17, no. 12, pp.~7981--7992, Dec. 2018.

\bibitem{Yuan2015}
Y. Yuan, C.-X. Wang, Y. He, M. M. Alwakeel, and H. Aggoune, ``3D wideband non-stationary geometry-based stochastic models for non-isotropic MIMO vehicle-to-vehicle channels,'' {\it IEEE Trans. Wireless Commun.}, vol. 14, no. 12, pp. 6883--6895, Dec. 2015.

\bibitem{SWu2018}
 S. Wu, C.-X. Wang, H. Aggoune, M. M. Alwakeel, and X. You, ``A general 3D non-stationary 5G wireless channel model,'' {\it Trans. Commun.}, vol. 66, no. 7, pp. 3065--3078, July 2018.

\bibitem{Yuan14}
Y. Yuan, C.-X. Wang, X. Cheng, B. Ai, and D. I. Laurenson, ``Novel 3D geometry-based stochastic models for non-isotropic MIMO vehicle-to-vehicle channels,'' {\it IEEE Trans. Wireless Commun.}, vol. 13, no. 1, pp.~298--309, Jan. 2014.

\bibitem{Wang09}
C.-X. Wang, X. Cheng, and D. I. Laurenson, ``Vehicle-to-vehicle channel modeling and measurements: recent advances and future challenges,'' {\it IEEE Commun. Mag.}, vol. 47, no. 11, pp. 96--103, Nov. 2009.

\bibitem{Dixit2012}
V. Dixit, S. Shukla, and M. Shukla, ``Performance analysis of optical IDMA system for indoor wireless channel model,'' in {\it Proc. IEEE CICN'12}, Mathura, India, Nov. 2012, pp. 382--386.

\bibitem{Maurer2005}
J. Maurer, T. Fuegen, and W. Wiesbeck, ``A ray-optical channel model for vehicular ad-hoc networks,'' in {\it Proc. IEEE EW'05}, Nicosia, Cyprus, 2005, pp. 1--7.

\bibitem{Barry1993}
J. R. Barry, J. M. Kahn, W. J. Krause, E. A. Lee, and D. G. Messerschmitt, ``Simulation of multipath impulse response for indoor wireless optical channels,'' {\it IEEE J. Sel. Areas Commun.}, vol. 11, no. 3, pp. 367--379, Apr.~1993.

\bibitem{Barry1994}
 J. R. Barry, {\it Wireless Infrared Communications}, New York: Springer, 1994.

\bibitem{European Commission}
D. Hynd, M. McCarthy, J. Carroll, M. Seidl, M. Edwards, C. Visvikis, M. Tress, N. Reed, and A. Stevens, ``Benefit and feasibility of a range of new technologies and unregulated measures in the fields of vehicle occupant safety and protection of vulnerable road users,'' European Commission, Brussels, Belgium, 2015.

\bibitem{Lindsey1997}
J. L. Lindsey, 2nd Ed., {\it Applied Illumination Engineering}, Lilburn: Fairmont Press, 1997.

\bibitem{Gfeller79} 
F. R. Gfeller and U. H. Bapst, ``Wireless in-house data communication via diffuse infrared radiation,'' {\it Proc. IEEE}, vol. 67, no. 11, pp. 1474--1486, Nov. 1979.

\bibitem{Ghassemlooy13_Book} 
Z. Ghassemlooy, W. Popoola, and S. Rajbhandari, 1st Ed., {\it Optical Wireless Communications: System and Channel Modelling with MATLAB}, New York: CRC press, 2013.

\bibitem{2D MIMO V2V}
X. Cheng, C.-X. Wang, D. I. Laurenson, S. Salous, and A. V. Vasilakos, ``An adaptive geometry-based stochastic model for non-isotropic MIMO mobile-to-mobile channels,'' {\it IEEE Trans. Wireless Commun.}, vol. 8, no. 9, pp. 4824--4835, Sept. 2009.

\bibitem{Zhuletter}
Q. Zhu, W. Li, C.-X. Wang, D. Xu, J. Bian, X. Chen, and W. Zhong, ``Temporal correlations for a non-stationary vehicle-to-vehicle channel model allowing velocity variations,'' {\it IEEE Commun. Lett.}, vol. 23, no. 7, pp. 1280--1284, July 2019.

\bibitem{Wangsurvey}
C.-X. Wang, J. Bian, J. Sun, W. Zhang, and M. Zhang, ``A survey of 5G channel measurements and models,'' {\it IEEE Commun. Surveys Tuts.}, vol. 20, no. 4, pp. 3142--3168, 4th Quart., 2018.

\bibitem{Elgala2009} 
H. Elgala, R. Mesleh, and H. Haas, ``Practical considerations for indoor wireless optical system implementation using OFDM,'' in {\it Proc. ConTEL'09}, Zagreb, Croatia, June 2009, pp. 25--29.

\bibitem{Jungnickel2002}
V. Jungnickel, V. Pohl, S. Nonnig, and C. V. Helmolt, ``A physical model of the wireless infrared communication channel,'' {\it IEEE J. Sel. Areas Commun.}, vol. 20, no. 3, pp. 631--640, Apr. 2002.

\bibitem{Mardia2000}  K. V. Mardia and P. E. Jupp, {\it Directional Statistics}. London: John Wiley \& Sons, 2000.

\bibitem{Bian2019}
J. Bian, C.-X. Wang, J. Huang, Y. Liu, J. Sun, M. Zhang, and H. Aggoune, ``A 3D wideband non-stationary multi-mobility model for vehicle-to-vehicle MIMO channels,'' {\it IEEE Access}, vol. 7, no. 1, pp. 32562--32577, Dec. 2019.

\bibitem{Ahmed dual}
A. Al-Kinani, C.-X. Wang, F. Haider, H. Haas, W. Zhang, and X. Cheng, ``Light and RF dual connectivity for the next generation cellular systems,'' in {\it Proc. IEEE ICCC'17}, Qingdao, China, Oct. 2017, pp. 1--6.

\bibitem{Lanewidth}
Transport Canberra and City Services. (Mar. 17, 2016). {Design standards for urban infrastructure}. Accessed: Mar. 30, 2020. [Online]. Available:
https://www.tccs.act.gov.au/.

\bibitem{Roadsidewidth}
Alta Planning and Design. (Jan. 2017). Urban, Rural and Suburban Complete Streets Design Manual. Cambridge, MA, USA. Accessed: Feb. 18, 2020. [Online]. Available:
https://www.northamptonma.gov/DocumentCenter/View/6668.

\bibitem{stoppingdistance}
Department for Transport. (Jan. 4, 2007). {Speed, Speed Limits and Stopping Distances}. Accessed: Mar. 31, 2020. [Online]. Available:
http://www.brake.org.uk/rsw/15-facts-a-resources/facts/1255-speed.

\bibitem{reflectivity}
M. Fabian, E. Lewis, T. Newe, and S. Lochmann, ``Optical fibre cavity for ring-down experiments with low coupling losses,'' {\it Meas. Sci. Technol.}, vol. 21, pp. 1--5,  July 2010.

\bibitem{concrete}
M. L. Marceau and M. G. VanGeem. (Aug. 2008). Solar Reflectance Values of Concrete. Illinois, USA. Accessed: Mar. 31, 2020. [Online]. Available:
http://www.cement.org.

\bibitem{Theory2017}
Z. Ghassemlooy, L. N. Alves, S. Zv$\acute{\mathrm{a}}$novec, and M.-A Khalighi, {\it Visible Light Communications Theory and Applications}, New York: CRC press, 2017.

\bibitem{Luo2014}
P. Luo, Z. Ghassemlooy, H. Le Minh, E. Bentley, A. Burton, and X. Tang, ``Fundamental analysis of a car to car visible light communication system,'' in {\it Proc. IEEE CSNDSP'14}, Manchester, UK, 2014, pp. 1011--1016.

\bibitem{Pätzold12}
 M. P{\"a}tzold, 2nd Ed., {\it Mobaile Radio Channels}, Chichester: John Wiley \& Sons, 2012.
\end{thebibliography}
\end{document}